\documentclass[manuscript,screen]{acmart}

\AtBeginDocument{%
  }

\setcopyright{none}
\copyrightyear{2026}
\acmYear{2026}
\acmDOI{XXXXXXX.XXXXXXX}

\usepackage{graphicx}%
\usepackage{multirow}%

\usepackage{amsmath,amssymb,amsfonts}%
\usepackage{amsthm}%
\usepackage{mathrsfs}%
\usepackage[title]{appendix}%
\usepackage{xcolor}%
\usepackage{textcomp}%
\usepackage{booktabs}%
\usepackage{tabularx}%
\usepackage{listings}%
\usepackage{tikz}%
\usepackage{float}%
\usetikzlibrary{calc,arrows.meta,positioning}
\begin{document}

\title{Spec-Driven Development for Agentic Software Engineering: Harnessing Human--Agent Teamwork}

\author{Jessica D\'iaz}
\email{yesica.diaz@upm.es}
\affiliation{%
  \institution{Universidad Polit\'ecnica de Madrid}
  \city{Madrid}
  \country{Spain}}

\author{Joaqu\'in Gayoso}
\email{joaquin.gayoso@upm.es}
\affiliation{%
  \institution{Universidad Polit\'ecnica de Madrid}
  \city{Madrid}
  \country{Spain}}

\author{Andrea Cimminio}
\email{andrea.cimminio@upm.es}
\affiliation{%
  \institution{Universidad Polit\'ecnica de Madrid}
  \city{Madrid}
  \country{Spain}}

\author{Jorge P\'erez}
\email{jorge.perez@upm.es}
\affiliation{%
  \institution{Universidad Polit\'ecnica de Madrid}
  \city{Madrid}
  \country{Spain}}

\renewcommand{\shortauthors}{D\'iaz et al.}

\begin{abstract}
\textbf{Context:} Software engineering is moving from AI-assisted practices like vibe coding, in which assistants accelerate the work of individual developers, towards Agentic Software Engineering (ASE) in which autonomous agents are delegated goal-level tasks. However, from the industrial domain a productivity paradox has been reported; as individual developers productivity increases the team-level throughput, review capacity, and stability degrades since the discipline that team-scale software engineering requires is neglected.
\textbf{Objective:} This paper aims to establish the conceptual and methodological foundations of Spec-Driven Development SDD as an enabling discipline for the operation of ASE at team scale and characterize the \emph{harness}, i.e., the set of technical and methodological mechanisms through which teams govern agent behavior.
\textbf{Method:} We conducted a conceptual analysis that, by necessity, draws predominantly on gray literature, i.e., the emerging vision and roadmap papers on ASE, together with practitioner reports, talks, and tooling. The reason is that peer-reviewed evidence and a shared academic--industrial vocabulary are not established in the field. 
\textbf{Results:} Using a comparative characterization of the paradigm progression as conceptual framing, the article presents (i) a socio-technical model of SDD in which specifications act as the contract substrate between humans and agents; (ii) an operational characterization of the harness, distinguishing the technical harness around the agent from the methodological harness around the team, and describing its mechanisms by means of worked examples; and (iii) a typology of five human--agent interaction patterns through which the human role is redefined.
\textbf{Conclusion:} As a result, authors conclude that SDD reconstitutes, in a specification-centric form, the contracts that vibe coding dissolves, i.e., accountability, verifiability, and transferability. Given the immaturity of the evidence base, this work is explicitly presented as a first step toward an academic--industrial consensus, rather than as a validated theory, and we outline a research agenda for its empirical validation in future.
\end{abstract}


\maketitle

\section{Introduction}\label{sec:introduction}

During these recent last years (from 2024 to 2026), autonomous coding agents (e.g., Claude Code, OpenAI Codex) moved from research demonstrations to production use in industry \citep{he2025llmagents,ahmed2025,amalfitano2026,li2026teammates,liu2026}. These agents are the most autonomous manifestation to date of Generative AI (GenAI), i.e., the family of foundation-model technologies that derive code, text, and other artifacts from natural-language input\footnote{It is worth distinguishing both terms from the outset, since GenAI had already entered software engineering practice in a passive form, that of the assistant that responds to a developer's prompt, whereas an agent employs the same underlying technology in an active form, i.e., it holds its own thread of control and decides which actions to perform and when. The distinction is not merely lexical, as the present paper will argue that it is precisely this shift from passive to active that reconfigures the socio-technical structure of the team.}. This short period of time has not been long enough for the vocabulary of the field has been established or agreed; terms such as \emph{vibe coding} \citep{karpathy2025vibe} and Agentic Software Engineering (\emph{SE 3.0 or ASE}) \citep{hassan2024se30} are being adopted, contested, and redefined almost simultaneously. This terminological instability reflects an instability of practice and hinders its efficient in adoption. At least three development models currently coexist in industry: (i) traditional and Agile development with light GenAI assistance, in which the structure of the Software Development Life Cycle (SDLC) remains unchanged; (ii) vibe coding, in which an individual developer maintains a free-form natural-language dialogue with an GenAI assistant and accepts, rejects, or refines its outputs in a tight loop; and (iii) ASE, in which agents are delegated goal-level tasks and humans assume the role of orchestrators and verifiers.

It is worth mentioning that these models are not points on a smooth continuum, since each of them imposes different demands on team structure, artifact discipline, and governance. A team that is competent in chat-assisted coding does not, by that fact alone, become competent in ASE. Moreover, increasing reliance on conversational and agent-assisted development practices may expose limitations in existing team coordination and governance mechanisms, a phenomenon that several industrial reports have begun to document. In fact, the empirical motivation of this paper is what several industrial reports describe as a productivity paradox. The Faros AI report, based on more than 10,000 developers in 1,255 teams, found that GenAI adoption was associated with a 21\% increase in individual task completion and a 98\% increase in merged pull requests (PRs), but also with a 91\% increase in PR review time and a 9\% increase in defects per developer \citep{farosai2025paradox}. The DORA reports describe a comparable tension over time: in 2024, higher AI adoption was associated with reduced delivery performance and greater team-level instability \citep{dora2024}, whereas the 2025 edition found the association with throughput reversed while instability persisted, and summarized the pattern by stating that AI does not fix a team but amplifies what is already there \citep{dora2025}. In the same vein, a randomized controlled trial conducted by METR found that experienced developers took 19\% longer to complete tasks in their own repositories while believing they had been about 20\% faster \citep{metr2025}, and the Stack Overflow 2025 survey identifies ``almost correct, but not quite'' outputs as the most frequent frustration with AI tools \citep{stackoverflow2025}. A further signal comes from the code itself, since across 211 million changed lines authored between 2020 and 2024 the share associated with refactoring fell from 25\% to below 10\%, while duplicated code blocks increased eightfold \citep{gitclear2025}. Hence, the degradation is observed not only in how fast teams deliver, but also in the maintainability of what they deliver.


\begin{figure}[t]
    \centering
    \includegraphics[width=0.8\textwidth]{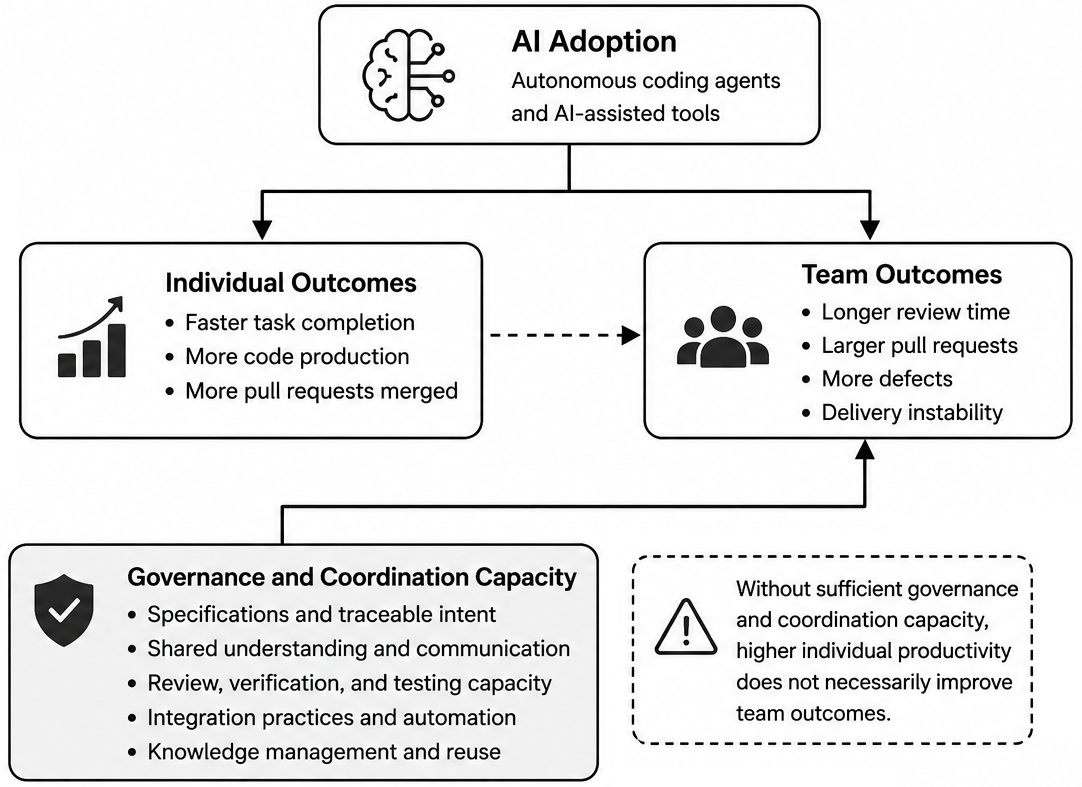}
    \caption{The productivity paradox and the role of governance in ASE.}
    \label{fig:productivity-paradox}
\end{figure}

Figure~\ref{fig:productivity-paradox} synthesizes this phenomenon. While GenAI adoption can substantially increase individual productivity, the resulting gains do not necessarily translate into improved team outcomes when governance, review, verification, and coordination capacities remain unchanged. 
Taken together, these data points suggest that the limiting factor is not model capability alone, but the absorption capacity of the surrounding socio-technical model. 
Hence, accelerating one phase of the SDLC (code production) does not accelerate the system as a whole if the downstream phases (review, verification, integration) retain their original capacity; in this case, the value created upstream is consumed by downstream bottlenecks. This effect is particularly relevant in modern code-review-based workflows, where review capacity is frequently the limiting resource rather than code production capacity \citep{bacchelli2013expectations}. This is, in essence, Amdahl's law \citep{amdahl1967} applied to the SDLC. In early Agile adoption a closely related pattern has been observed before; the acceleration introduced by short Scrum sprints often overwhelmed downstream system and operations teams, which were unable to absorb the increased delivery pace and consequently became bottlenecks. The phenomenon resembles the optimization of local subsystems at the expense of global flow, a challenge previously identified in lean software development and systems thinking approaches. This tension was one of the drivers behind the emergence of DevOps as a practice to rebalance end-to-end system capacity, as it has been analyzed in authors previous work \citep{diaz2021many}. The intermediate practice of vibe coding provides perhaps the clearest illustration of this paradox, as it emphasizes rapid individual code production while often relying on conversational interactions that leave limited traceability, rationale preservation, or reusable specification artifacts.

In this article, the authors delve about the transition from GenAI-augmented SE to ASE is principally a socio-technical reconfiguration rather than a technological event, and hypothesize that \textbf{Spec-Driven Development} (\textbf{SDD}) is the discipline that makes this transition feasible. More generally, the centrality of specifications is not new in software engineering. Requirements Engineering has long emphasized the role of specifications as communication and validation artifacts between stakeholders and developers \citep{nuseibeh2000requirements,sommerville2011requirements}. What changes in the agentic paradigm is that specifications become operational artifacts consumed directly by autonomous agents. Under SDD, specifications become the canonical, version-controlled contract between humans and agents, from which code, tests, documentation, and infrastructure are derived with a clear trace of their origin. In addition, the authors also argue that the adoption of SDD must be understood together with the notion of \textbf{\emph{harness}}, i.e., the set of mechanisms, both technical and methodological, that surround the agent and the team, and that help translate the agent's capabilities into meaningful outcomes for the team. The harness concept has recently emerged in the practitioner literature as the infrastructure wrapping a model (Agent = Model + Harness) \citep{fowler2026harness,langchain2026harness}; in this paper, the harness is extended to the team level and connected to SDD. The progression from Agile to ASE and the limitations of vibe coding therefore provide the conceptual framing for the paper. Against this framing, authors propose two research questions:

\begin{itemize}
\item \textbf{RQ1.} Which mechanisms constitute the harness through which teams govern agent behavior, and what is the role of specifications within it?
\item \textbf{RQ2.} Which interaction patterns enable effective human--agent collaboration under SDD, and how must human roles be redefined?
\end{itemize}

Before proceeding, authors need to highlight a methodological caveat on how the contributions of this paper should be interpreted. The phenomenon under analysis is barely two years old, although it would be inaccurate to claim that it has escaped academic attention altogether. Systematic reviews and roadmap articles have already mapped the landscape of GenAI-augmented and agentic SE in a traceable manner \citep{he2025llmagents,wang2025agents,hassan2025agentic,amalfitano2026,liu2026,hoda2026agentic}. What is missing is more specific than a general absence of evidence. The two constructs on which this paper rests, i.e., Spec-Driven Development understood as a team-level discipline and the harness understood as the set of mechanisms through which teams govern agent behavior, have not yet been the object of academic treatment. Both were named in practitioner discourse and are still being refined there, and, to the best of our knowledge, no peer-reviewed study has defined them, delimited their scope, or measured their effects.
Considering that the available knowledge on these two constructs lives in \emph{gray literature}: vision papers (many of them preprints), conference talks, practitioner blog posts, and open-source tooling whose documentation reflects operational practices acquired through great effort. As a result, authors deliberately and explicitly have built the reported analysis on this gray literature, and it is treated as such, i.e., as the best available evidence in an immature field rather than as settled science. We do not regard this as a weakness to be hidden, but as the defining condition of any honest contribution at this moment: the academic consensus that would allow us to do otherwise does not exist yet, and someone has to take the first step toward building it. The present article is intended as a structured synthesis that may help establishing a shared conceptual foundation for future empirical work. 
The contributions of this paper are the following:

\begin{itemize}

\item \textbf{C1.} A socio-technical model of SDD that positions specifications as the contract substrate of human--agent collaboration (Section~\ref{sec:sdd}).

\item \textbf{C2.} An operational characterization of the harness for ASE, distinguishing the technical harness from the methodological harness and describing its constituent mechanisms with worked examples (Section~\ref{sec:harness})

\item \textbf{C3.} A typology of five human--agent interaction patterns through which the human role is redefined under SDD (Section~\ref{sec:patterns}).
\end{itemize}

The structure of the paper is as follows. Section~\ref{sec:methodology} is devoted to the research methodology applied
in this work. Section~\ref{sec:background} traces the paradigm progression that motivates the framework. Section~\ref{sec:sdd} and Section~\ref{sec:harness} establish SDD as the enabling discipline and characterize the harness (RQ1). Section~\ref{sec:patterns} develops the interaction patterns and the redefinition of human roles (RQ2). Section~\ref{sec:traceability} reports the traceability of the synthesis. Section~\ref{sec:discussion} discusses benefits, risks, and threats to validity. Section~\ref{sec:agenda} outlines a research agenda, including the study of organizational adoption pathways, which we deliberately leave as future work. Finally, Section~\ref{sec:conclusion} draws the main conclusions of our work.




\section{Background: From Agile to Agentic}\label{sec:background}

Before making claims about what ASE requires from teams, let us first establish what each prior paradigm requires and what each transition changes. To this end, the section compares four development paradigms along five socio-technical dimensions: the unit of work, the primary artifact, the locus of cognition, the structure of accountability, and the team topology (the latter of which was analyzed in depth in our previous work \cite{lopez2021}).

\begin{figure}[t]
\centering

\begin{tikzpicture}[
    node distance=1.45cm,
    paradigm/.style={
        draw,
        rounded corners,
        align=center,
        minimum width=5.4cm,
        minimum height=1.1cm
    },
    sidebox/.style={
        draw,
        dashed,
        rounded corners,
        align=left,
        font=\small,
        inner sep=5pt,
        text width=4.1cm
    },
    arrow/.style={-{Stealth[length=3mm]}, thick},
    gaparrow/.style={-{Stealth[length=3mm]}, thick, dashed}
]

\node[paradigm] (agile) {
\textbf{Agile \& DevOps}\\
Human-authored software\\
Individual accountability
};

\node[paradigm, below=of agile] (ai) {
\textbf{GenAI-Augmented SE 2.0}\\
Human + assistant\\
Human accountability preserved
};

\node[paradigm, below=of ai] (vibe) {
\textbf{Vibe Coding}\\
Conversational development\\
Weak governance and traceability
};

\node[paradigm, below=2.2cm of vibe] (agentic) {
\textbf{ASE + SDD}\\
Specifications as primary artifact\\
Contractual accountability
};

\draw[arrow] (agile) -- (ai);
\draw[arrow] (ai) -- (vibe);
\draw[gaparrow] (vibe) -- (agentic);

\node[
    sidebox,
    right=0.8cm of $(vibe)!0.5!(agentic)$
] (gap) {
\textbf{Governance gap}\\
Loss of traceability\\
Loss of auditability\\
Loss of transferability
};

\node[
    sidebox,
    right=1.2cm of ai
] (dims) {
\textbf{Socio-technical dimensions}\\[1mm]
$\bullet$ Unit of work\\
$\bullet$ Primary artifact\\
$\bullet$ Locus of cognition\\
$\bullet$ Accountability\\
$\bullet$ Team topology
};

\end{tikzpicture}

\caption{
Evolution of software engineering paradigms toward ASE.
Vibe coding introduces a governance gap caused by the loss of traceability,
auditability, and transferability. SDD restores coordination through
specifications as the primary artifact of human--agent collaboration.
}
\label{fig:paradigm-evolution}

\end{figure}
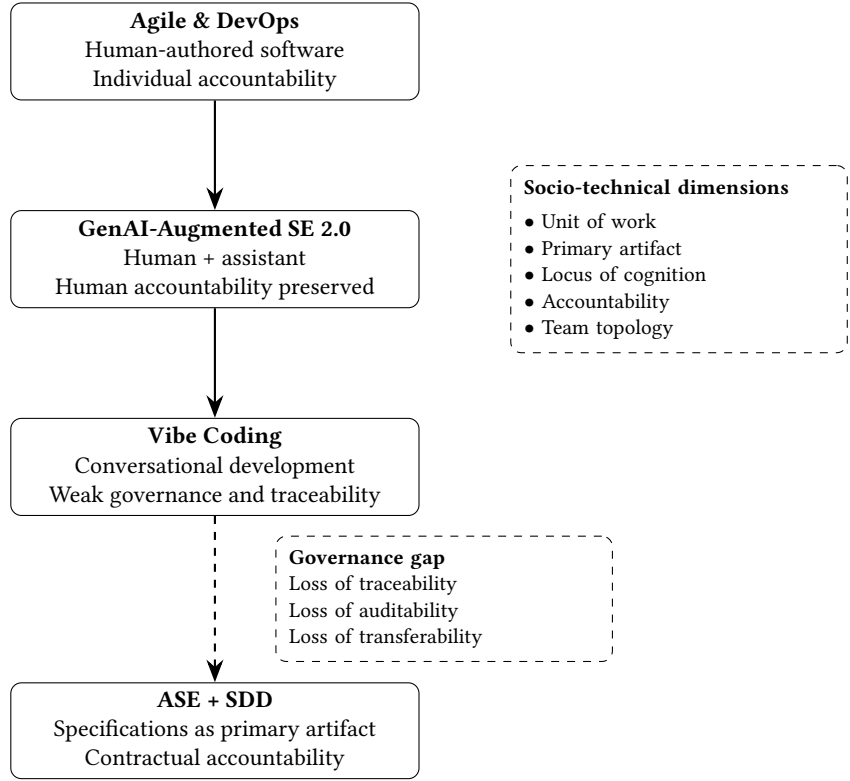

Figure~\ref{fig:paradigm-evolution} summarizes the progression analyzed in this section. While the transition from Agile to GenAI-augmented SE mainly affects the execution of individual tasks, the shift towards vibe coding and ASE fundamentally changes the socio-technical structure of software development teams.

\subsection{Agile \& DevOps}\label{sec:agile}

Agile is a paradigm been consolidated through Scrum and its variants \citep{hoda2018agile}. It assumes a team of human developers supported by IDEs, version control, automated builds, and code review platforms. In this paradigm, the unit of work is the user story; the primary artifact is working software; the locus of cognition is the individual developer (with collaboration through pairing and review); accountability is individual-to-team, i.e., the developer who wrote the code answers for it; and the team topology is small, cross-functional, and persistent, as we analyzed in our previous work \citep{lopez2021}. It is worth mentioning that Agile systematically under-weights documentation relative to working software \citep{Beck:2001}, a debt that becomes visible mainly at hand-off and onboarding. Moreover, its review practices were built around the assumption that code is authored by humans and evolves at a human pace. Hence, Agile does not anticipate a team member that is not human, that produces artifacts faster than any human can review them, and that has no professional accountability. Despite these limitations, the Agile paradigm succeeds because it externalizes coordination through shared artifacts, stable team structures, explicit review practices, and a common understanding of responsibility. These mechanisms are often taken for granted precisely because they have been refined through decades of industrial adoption. As long as software artifacts are primarily authored by humans, they provide sufficient traceability for review, onboarding, and accountability. The emergence of autonomous agents challenges these assumptions by increasing the volume, speed, and opacity of software production beyond the levels for which Agile practices were originally designed.

\subsection{GenAI-augmented software engineering (SE 2.0)}\label{sec:se2}

GenAI-augmented SE has emerged between 2022 and 2024, it integrates assistants such as GitHub Copilot or Cursor into the developer's IDE \citep{hassan2024se30}. Structurally, the change that introduces is narrow: the unit of work, the primary artifact, accountability, and topology remain those established in the Agile \& DevOps paradigms. Only specific sub-activities, such as generating boilerplate code, recalling API syntax, or translating patterns between languages, are accelerated by AI agents. Nevertheless, two problems emerge in the GenAI-augmented SE paradigm. The first one is the ``almost correct'' problem: code that is almost correct is more expensive to repair than code that is obviously incorrect, because the failure mode resists debugging \citep{stackoverflow2025}. The second one is the context problem: an IDE-integrated assistant has access to a few open files and the immediate prompt, but it lacks the cross-cutting knowledge of conventions and prior decisions that a human developer accumulates over months \citep{qodo2025}.

\subsection{Vibe coding}\label{sec:vibe}

The term vibe coding was coined by Karpathy in early 2025 for naming the practice of working with an GenAI assistant in continuous natural-language dialogue, accepting the outputs that feel right and iterating until subjective satisfaction is reached \citep{karpathy2025vibe}. Unlike GenAI-augmented SE, where the developer remains the primary producer of software artifacts, vibe coding shifts substantial portions of problem solving and implementation to the conversational interaction itself. In this practice, the unit of work is expanded from the user story to the whole feature, or even the whole system; the primary artifact is code that is increasingly AI-authored in its initial form; the locus of cognition shifts toward the conversational interface; accountability becomes ambiguous (the developer signs the PR but did not write the code); and the team topology fractures functionally as the individual rates of code production diverge.

Vibe coding can be locally effective, i.e., a capable developer can produce in hours what previously took days. However, our analysis identifies five recurring limitations that make it difficult to scale as a team-level practice unless additional governance mechanisms are introduced:

\begin{enumerate}
\item \textbf{Non-reproducibility.} The same prompt in the same context can produce different code. Without a stable specification, the team cannot answer the question ``why is this code here?'' except by reconstructing a conversation that is rarely preserved.
\item \textbf{Non-auditability.} The reasoning that produced a change is buried in a transient chat log, so the reviewer has no artifact to assess other than the code itself.
\item \textbf{Non-transferability.} A developer who leaves the team takes the conversational context with them. Therefore, onboarding a successor means re-discovering, by inference from code, what was once explicit in dialogue.
\item \textbf{Review saturation.} \citet{li2026teammates,hassan2025agentic} report that a majority of agent-generated PRs experience long delays or remain unreviewed; the reviewer cannot recover the intent and must reconstruct it from the diff.
\item \textbf{Noise scaling.} When several developers use AI to generate and refine code independently, the codebase diverges across naming, structure, and architectural choices, which increases inconsistency and fragmentation.
\end{enumerate}

It is worth  to mention that these are failures of methodology and not of practitioners doing. Vibe coding is the natural exploratory phase of a new tool before disciplines exist to guide and structure its use. In the case of vibe coding, the productivity paradox can be interpreted as one observable symptom of this lack of team-level discipline.

\subsection{Agentic software engineering (ASE or SE 3.0)}\label{sec:se3}

The agentic paradigm treats GenAI agents as actors capable of executing multi-step, goal-level tasks with relative autonomy \citep{hassan2025agentic,roychoudhury2025trust}. In this paradigm, the unit of work is the technical goal (e.g., ``add rate limiting to the orders API''). The primary artifact is the specification from which code, tests, documentation, and infrastructure are derived. The cognition is redistributed between humans (intent, oversight, strategy) and agents (execution, parallel exploration); accountability becomes contractual, i.e., relative to the specification; and the team topology becomes N-to-N, with multiple humans collaborating with multiple agents.

\citet{hassan2025agentic} capture this paradigm through the duality of SE for Humans (SE4H) and SE for Agents (SE4A), connected by versioned artifacts (briefing documents, workflow definitions, codified normative rules, consultation requests, merge-readiness evidence). Note that each of these artifacts is, in essence, a specification of a different kind. At this point SDD becomes relevant: ASE without a specification discipline risks reproducing the same weaknesses of vibe coding at a larger scale: faster execution, broader scope, and weaker coordination.

\subsection{Comparative synthesis}\label{sec:synthesis}

Table~\ref{tab:paradigms} summarizes the four development paradigms. From this comparison, two observations deserve explicit mention. First, the transition from SE 2.0 to vibe coding is regressive on the dimensions of accountability, verification, and onboarding, since it gains individual velocity at the cost of the team-scale mechanisms that made the previous paradigm sustainable. Second, the transition from vibe coding to ASE under SDD recovers those dimensions in a transformed shape: accountability becomes contractual against the specification, verification audits evidence rather than reconstructing intent, and onboarding reads specifications rather than code. This is not a return to pre-AI Agile \& DevOps, but a forward step that exploits AI agency at scale while reconstituting the contracts that make teamwork possible.

\begin{table}[ht]
\caption{Comparative socio-technical structure of four software engineering paradigms.}\label{tab:paradigms}
\footnotesize
\begin{tabular*}{\textwidth}{@{\extracolsep\fill}p{2.0cm}p{2.4cm}p{2.4cm}p{2.2cm}p{2.4cm}@{}}
\toprule
\textbf{Dimension} & \textbf{Agile \& DevOps} & \textbf{AI-Augmented (SE 2.0)} & \textbf{Vibe Coding} & \textbf{ASE with SDD (SE 3.0)} \\
\toprule
Unit of work & User story & User story & Feature / micro-app & Technical goal \\
\midrule
Primary artifact & Working software & Working software (assisted) & Code (AI-authored) & Specification \\
\midrule
Locus of cognition & Developer & Developer + assistant & Conversational interface & Distributed (humans + agents) \\
\midrule
Accountability & Individual-to-team & Individual-to-team & Ambiguous & Contractual (via specification) \\
\midrule
Team topology & Cross-functional squad & Cross-functional squad & Individual-centered & N-to-N human--agent network \\
\bottomrule
\end{tabular*}
\end{table}

\subsection{Related work}\label{sec:related}

Our work builds on three streams of related work. The first one is the emerging literature on ASE, in particular the vision and roadmap papers by \citet{hassan2024se30,hassan2025agentic}, \citet{hoda2026agentic}, and \citet{roychoudhury2025trust}, as well as surveys of LLM-based agents for SE \citep{he2025llmagents,wang2025agents,ahmed2025,liu2026}. Within this stream, the roadmap of \citet{amalfitano2026} deserves separate mention, both for its scope and for its methodological rigor: following a design-science process with three cycles and rapid literature reviews, it structures GenAI augmentation along two dimensions, i.e., what is augmented (process or product) and how autonomous the augmentation is (passive or active), and derives four forms from their intersection, one of which, the actively autonomous augmentation of the SE process, is precisely the territory of the present paper. Their analysis is descriptive since it characterizes what the technology enhances, reverses, retrieves, and renders obsolete, and it culminates in a catalog of research challenges. Our contribution with respect to this stream is complementary rather than competing, i.e., where they map the landscape across the four forms, we go into depth on one of them and propose the discipline, the artifacts, and the team-level mechanisms through which that form can be operated. Our contribution is, therefore, to frame the transition as a socio-technical problem and to place specifications at its center.

The second stream is the long tradition of specification-centric approaches in SE, including Model-Driven Engineering, Domain-Driven Design \citep{evans2003ddd}, and executable specification practices such as Behavior-Driven Development, recently taken up in agentic settings by \citet{zhang2025}. SDD inherits from these traditions the principle of an authoritative abstraction, but it differs in that the transformation engine is a stochastic agent rather than a deterministic generator, which changes the role of verification (see Section~\ref{sec:harness}). 

The third stream is practitioner knowledge, which in this field is unusually rich and, at the same time, unusually informal: the discourse on the \emph{agent harness} \citep{fowler2026harness,langchain2026harness}, the empirical telemetry on GenAI-assisted development outcomes \citep{farosai2025paradox,dora2024,metr2025}, and the documented practice of open-source agentic-development toolchains. Authors draw on this third stream knowingly. As noted in Section~\ref{sec:introduction}, much of the evidence in this area is gray literature whose methodologies and definitions vary and are not peer-reviewed; rather than waiting for an academic consensus that has not formed yet, we synthesize this body of practice into a structured account that subsequent empirical work can test. Authors see this synthesis itself, i.e., the act of giving a contested and fast-moving practice a coherent conceptual frame, as part of the contribution.

Table~\ref{tab:sdd-related} clarifies this positioning. The novelty of SDD is not the use of specifications per se, since specification-centric approaches have existed for decades. Rather, SDD combines specifications with a fundamentally different execution model based on autonomous agents and a redefinition of the human role from artifact producer to orchestrator and verifier. In this context, specifications cease to be merely design or validation artifacts and become operational contracts that coordinate, constrain, and govern agent behavior throughout the software life-cycle.

\begin{table}[t]
\caption{Positioning SDD with respect to specification-centric approaches.}
\label{tab:sdd-related}
\footnotesize
\begin{tabular*}{\textwidth}{@{\extracolsep\fill}p{1.4cm}p{2.4cm}p{2.7cm}p{2.3cm}p{3.6cm}@{}}
\toprule
\textbf{Approach} &
\textbf{Primary abstraction} &
\textbf{Execution mechanism} &
\textbf{Human role} &
\textbf{Role of specifications} \\
\midrule

MDE &
Models &
Deterministic transformations &
Model author &
Generate or synchronize artifacts \\

BDD &
Scenarios &
Automated tests &
Requirement author &
Validate expected behavior \\

DDD &
Domain model &
Human implementation &
Domain expert &
Align software with domain concepts \\

SDD &
Specifications &
Autonomous stochastic agents &
Orchestrator and verifier &
Coordinate and govern human--agent collaboration \\

\bottomrule
\end{tabular*}
\end{table}

\section{Research Methodology}\label{sec:methodology}

This research is based mainly on the constructivist model as an underlying philosophy \cite{Easterbrook:2008}. Constructivism or interpretivism states that scientific knowledge cannot be separated from its human context. It also states that a phenomenon can be fully understood by considering the participants' perspectives and context. Consequently, this study is defined as a qualitative research  since its focus lies in understanding, structuring, and refining the semantic meaning of the agentic model and human-agent interaction (rather than on the analysis of dependent and independent variables and their quantification). This methodological design is supported operationally by a Multivocal Literature Review (MLR). This method allows for a structured synthesis that integrates emerging academic perspectives with the wealth of evidence available in the gray literature. We reviewed blog posts, videos and white papers in addition to the formal literature (e.g., journal and conference papers) according to the process defined by \cite{GAROUSI:2019}:  search process; source selection; study quality assessment; data extraction; and data synthesis. The discussion of threats to validity is approached from this qualitative perspective in Section~\ref{sec:threats}.

\subsection{Research Questions (RQs)}
In order to organise the extraction, synthesis and analysis of the evidence identified, this multivocal review is structured around two main research questions (RQs)
\begin{itemize}
\item \textbf{RQ1.} Which mechanisms constitute the harness through which teams govern agent behavior, and what is the role of specifications within it?
\item \textbf{RQ2.} Which interaction patterns enable effective human--agent collaboration under SDD, and how must human roles be redefined?
\end{itemize}

\subsection{Search process}
Due to the scarce and fragmented literature on the scope of this study, an exploratory search was performed targeting both traditional academic databases and grey literature sources

\paragraph{Search in Formal Academic Sources}
The main academic databases for Software Engineering and Artificial Intelligence—IEEE Xplore, ACM Digital Library, Scopus, and arXiv—were queried. The following search query was applied to the full text of the documents: (''Agentic Software Engineering´´). A broad search string was intentionally used due to the novelty of the field.

\paragraph{Grey Literature Search}
The search was partitioned into the following domains:
\begin{itemize}
\item	Technical Industry Reports: Works from organizations such as DORA / Google Cloud, Faros AI, METR, GitClear, Qodo, and Stack Overflow.
\item	Technical Publications and Whitepapers: Articles issued by specialized blogs and technical leaders, such as Martin Fowler and LangChain Research.
\item	Pioneering Technical Content Creators: Key voices in the conceptual discussion of SE 3.0, such as Andrej Karpathy on X/Twitter and MoureDev/Gentleman on YouTube.
\end{itemize}

\subsection{Selection Criteria}
Tabla~\ref{tab:inclusion-exclusion-criteria} summarizes the inclusion and exclusion criteria applied to the search results. The inclusion criteria (IC) were designed to capture works that explicitly address the agentic paradigm and provide empirical or methodological insights relevant to SDD. The exclusion criteria (EC) filter out works that focus solely on traditional code assistants, promotional content, or discussions that do not contribute to the understanding of ASE and SDD.

\begin{table}[ht]
\centering
\caption{Inclusion and exclusion criteria}
\label{tab:inclusion-exclusion-criteria}
\begin{tabularx}{\textwidth}{|X|X|}
\hline
\textbf{Inclusion Criteria (IC)} & \textbf{Exclusion Criteria (EC)} \\
\hline

\textbf{IC1:} Academic works or industry reports explicitly focused on the 
agentic paradigm (SE 3.0).
&
\textbf{EC1:} Publications exclusively focused on traditional code assistants 
(SE 2.0 / first-generation Copilot-style IDE autocomplete without agentic autonomy).
\\
\hline

\textbf{IC2:} Studies reporting empirical metrics or observational data on 
team-level productivity, code refinement rate, or deliverable dynamics.
&
\textbf{EC2:} Promotional articles, press releases, or marketing literature 
without transparent methodological data.
\\
\hline

\textbf{IC3:} Formal proposals on technical harnesses (agent harness), 
executable specifications, or interaction patterns.
&
\textbf{EC3:} Discussion posts on channels such as X or YouTube whose content 
relates to AI as an assistant.
\\
\hline

\end{tabularx}
\end{table}

\subsection{Selection Process}
The search and selection process (based on the IC/EC) yielded the following results, categorised by source:
\begin{itemize}
	\item	IEEE Xplore: 15 papers, of which 12 were included and 3 were excluded
	\item	ACM Digital Library: 21 papers, of which 16 were included and 5 were excluded
	\item	Scopus: 36 papers, of which 22 were included and 14 were excluded
	\item	arXiv: 48 papers, of which 44 were included and 4 were excluded
\end{itemize}

A total of 9 duplicates were identified amongst the papers. The full set of these papers can be found in the replication package that accompanies this article and is deposited in Zenodo under DOI \url{https://doi.org/10.5281/zenodo.22151221}. The papers are listed by source (those marked with a strikethrough are the duplicates identified).

\subsection{Study quality assessment}
For traditional academic sources, the impact of publication in the JCR index or texts endorsed by the scientific community has been considered. For grey literature (GL), some of the criteria outlined by Garousi et al. (2019, Table 7) have been used, including: i) the reputation of the information provider (e.g. DORA or METR); ii) date/novelty (as in the case of the 2026 studies); and iii) objectivity (such as the Faros AI report on AI productivity, based on data from more than 10,000 developers in 1,255 teams).

\subsection{Data extraction and data synthesis}
For each included source, the authors extracted evidence fragments related to team-level coordination, agent governance, specification use, review and verification practices, persistent knowledge, and human involvement in agentic workflows. These fragments were first assigned descriptive codes close to the source terminology. Codes were then compared across academic and gray-literature sources and grouped into higher-level categories through thematic synthesis. Categories that described recurrent governance mechanisms were consolidated into the harness mechanisms discussed in Section~\ref{sec:harness}; categories that described recurrent interaction moves between humans and agents were consolidated into the five interaction patterns discussed in Section~\ref{sec:patterns}. The synthesis was therefore interpretive rather than statistical: the patterns are conceptual constructs derived from repeated evidence across the reviewed sources, not empirically validated frequencies.

The resulting traceability from source evidence to extracted concepts, synthesized categories, and mechanisms or patterns is reported after the presentation of the SDD mechanisms and interaction patterns, in Section~\ref{sec:traceability}. The complete extraction matrix and source-selection material are provided in the replication package, which is deposited in Zenodo under DOI \url{https://doi.org/10.5281/zenodo.22151221}.

The analysis and synthesis of the data are set out in detail in the following sections. The section `From Agile to Agentic' provides the conceptual framing for both RQs. The section `Spec-Driven Development for Agentic SE' and the section `Harnessing Agentic Capability: The Technical and Methodological Harness' address RQ1. Finally, the section `Human--Agent Interaction Patterns and the Redefinition of Human Roles' addresses RQ2.

\section{Spec-Driven Development for Agentic SE}\label{sec:sdd}

This section examines how SDD can provide the team-level discipline required for ASE to operate at scale. Our presentation is socio-technical rather than formal, i.e., SDD is treated as a set of team-level commitments about how work flows, how artifacts are governed, and how humans and agents contract with one another.

\subsection{Operational definition and common misunderstandings}\label{sec:definition}

Authors define SDD operationally as follows. A team operates under SDD when:

\begin{enumerate}
\item Every non-trivial planned change to the system\footnote{By non-trivial planned change we refer to changes that affect behavior, architecture, interfaces, data, tests, infrastructure, or team conventions; purely mechanical edits, such as formatting corrections or typo fixes, may be handled without a dedicated specification.} originates from a written specification that the team treats as the contract for the change.
\item The specification is version-controlled alongside the code, with the same review, audit, and rollback discipline.
\item Artifacts (code, tests, documentation, infrastructure) are derived from the specification by humans, agents, or both, with explicit provenance and traceability back to the originating specification.
\item When specification and artifact disagree, the specification is the source of truth, and the disagreement is resolved by re-deriving the artifact or amending the specification, never by silent edits.
\end{enumerate}

These four commitments are simple to state, but they are demanding to operate. 
The emphasis on provenance, traceability, and authoritative specifications is not unique to SDD. Rather, SDD extends established principles from Requirements Engineering to human--agent collaboration, making specifications the anchor through which generated artifacts can be related to their originating intent and against which implementation and validation activities can be assessed \citep{gotel1994analysis,cleland2002traceability,nuseibeh2000requirements,sommerville2011requirements}.

In addition, a practical observation deserves to be stated as a first-class commitment: specifications exist at two levels, and confusing them is a frequent adoption failure.

\begin{itemize}
\item \textbf{System specifications} describe the durable world in which the project lives, i.e., architecture, conventions, domain models, patterns, and the normative rules the team expects every change to respect. They are typically owned by technical leads and architects, they change with architectural decisions rather than with feature work, and they are increasingly \emph{materialized} as machine-readable rule files that agents load at the start of every session (e.g., \texttt{agent.md}, \texttt{CLAUDE.md}, \texttt{.cursorrules}, \texttt{AGENTS.md}, or equivalent). We treat the rule file not as an incidental format detail, but as the canonical, version-controlled artifact in which the system specification physically lives.
\item \textbf{Feature specifications} describe a specific change within that world, i.e., what is being built now, with which acceptance criteria and constraints. They are owned by the developer responsible for the feature, created per task, and archived once delivered.
\end{itemize}

The distinction is analogous to the separation between organizational policy and operational work instructions. System specifications define the stable constraints under which all development takes place, whereas feature specifications define the objectives and acceptance criteria of a particular change. Both levels must be present in the agent's context. When only the feature specification is loaded, the agent produces code that is functionally correct but architecturally inconsistent; conversely, when only the system specification is loaded, the agent cannot ground the current task. Conflating the two causes either excessive rigidity (when feature-specific decisions are embedded into the system specification) or architectural drift (when global constraints are repeatedly redefined within feature specifications).

\begin{figure}[!htbp]
    \centering
    \includegraphics[width=\textwidth]{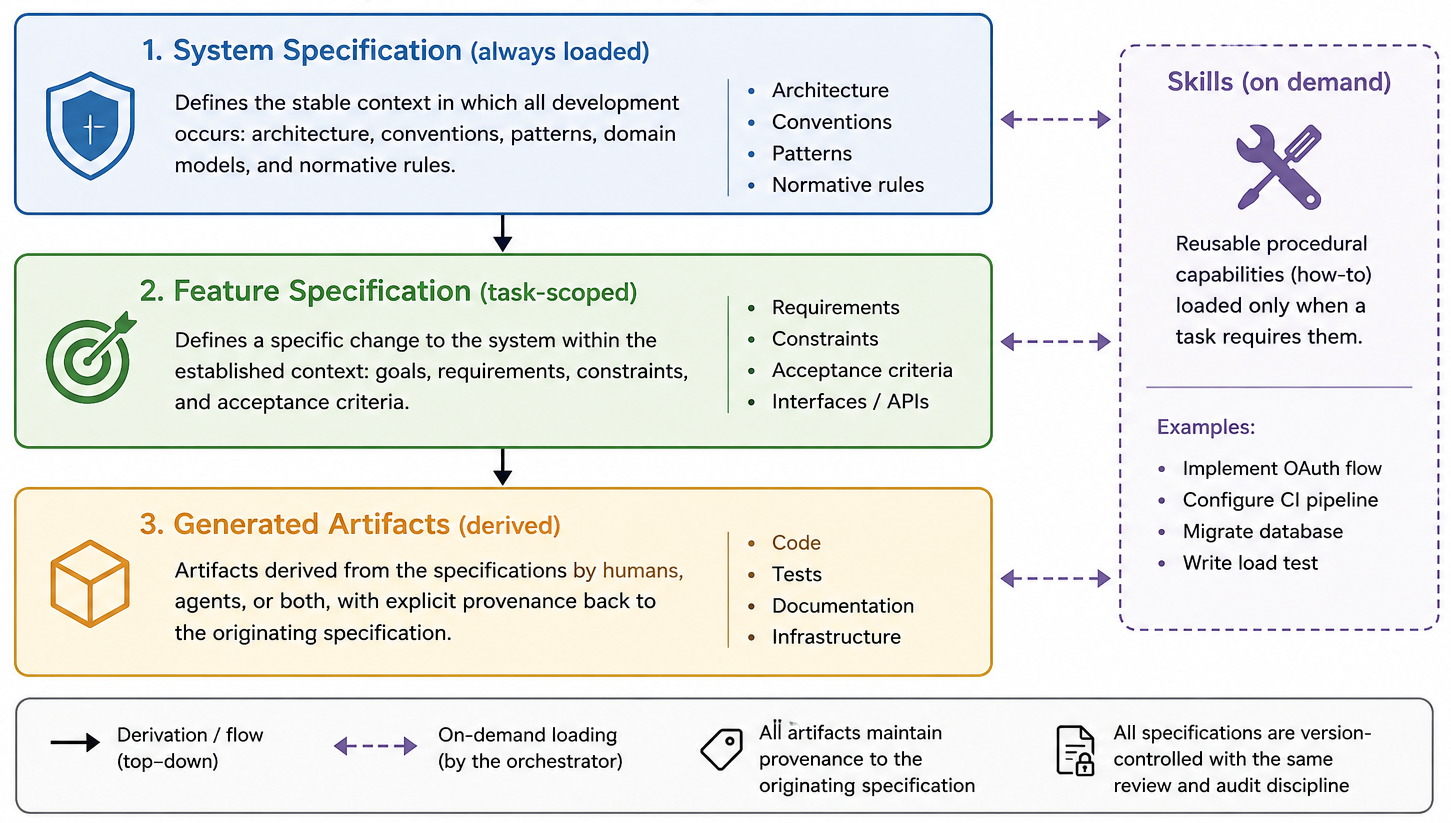}
    \caption{
    Specification hierarchy in SDD. System specifications provide the stable normative context of the project; feature specifications define task-specific changes; artifacts are derived from specifications; and skills provide reusable procedural capabilities loaded on demand.
    }
    \label{fig:sdd-hierarchy}
\end{figure}

A third kind of artifact must be distinguished from both, since conflating them is itself a common adoption error. \emph{Skills} (e.g., \texttt{SKILL.md} files indexed in a skill registry) are reusable, \emph{on-demand} capabilities, i.e., procedural know-how that the orchestrator loads only when a task triggers it, and deliberately not otherwise, so as not to pollute the agent's context with rules that do not apply to the current work. Hence, a skill is procedural (``how to do X'') and selectively loaded, whereas a system specification is normative (``what every change must respect'') and always active. The two complement each other, but they occupy different roles in the harness: skills are a context-engineering instrument (Section~\ref{sec:context}), whereas the normative content of the system specification is the mechanism we develop in Section~\ref{sec:normative}. This separation is important because it controls both context quality and governance. If procedural knowledge is placed in the system specification, the agent's always-loaded context becomes unnecessarily large and harder to maintain. Conversely, if normative rules are encoded as skills, mandatory constraints may only be applied when the corresponding skill happens to be loaded. SDD therefore separates stable norms, task-specific intent, and reusable procedures into different artifact classes.

Other common misunderstanding is believe that SDD entails ``more documentation''. The distinction is both temporal and functional. Documenting code that already exists is retrospective, i.e., it serves understanding after the fact and is intrinsically lagging. Specifying behavior before implementation is, on the contrary, an act of design: edge cases are discovered before they reach production, expectations are aligned before the PR is opened, and design decisions are made when they are still cheap to change. In an agentic model this distinction is amplified, because the agent consumes the specification at execution time. Thus, a vague specification is not merely poor documentation; it is poor input that produces poor output at scale.

Finally, other misunderstanding associates SDD with a waterfall process in which an exhaustive specification precedes all implementation. SDD in practice is iterative. A typical cycle, observable in current tooling (e.g., the proposal--apply--archive cycle of OpenSpec-style workflows), proceeds as follows: in the proposal phase, a lightweight specification is enriched iteratively, surfacing ambiguities and edge cases; in the apply phase, the agent derives artifacts without improvising outside the specification, and suspends to consult when it encounters ambiguity. Finally, in the archive phase, the verified specification is committed, and subsequent changes generate new specifications that reference it. Iteration is therefore preserved, but it happens over specifications first and code second.

\subsection{Specifications as the contract substrate for agentic SE}\label{sec:substrate}


The need for specifications in ASE emerges from a chain of dependencies. Specifications are not valuable in isolation; rather, they enable contracts, contracts enable coordination, and coordination enables trust at scales where direct human inspection is no longer feasible. The following three problems can therefore be understood as successive manifestations of the same underlying issue, which motivates the need for an SDD-like discipline in ASE.

\paragraph{The contract problem} An agent receiving a vague natural-language ticket will produce plausible code that is not what was wanted, and the cost of the mismatch is paid in review, rework, or production defects. The agentic model industrializes this cost: a team may receive many candidate PRs in parallel from multiple agents \citep{hassan2025agentic}, and reviewers cannot reconstruct intent across that volume. In this context, the specification is the contract that scopes what the agent must produce.

\paragraph{The coordination problem} Multi-agent teams collaborate through shared artifacts. If the shared artifact is the code, agents will write conflicting code. When is the specification, agents have a substrate over which decomposition, escalation, and merging become tractable.

\paragraph{The trust problem} Humans cannot establish trust in agent output by inspecting code line by line at agent-scale volumes. Trust must be established by reviewing evidence of conformance to a contract. Note that evidence only makes sense relative to a specification; without one, ``evidence'' has no referent.

\begin{figure}[t]
    \centering
    \includegraphics[width=\textwidth]{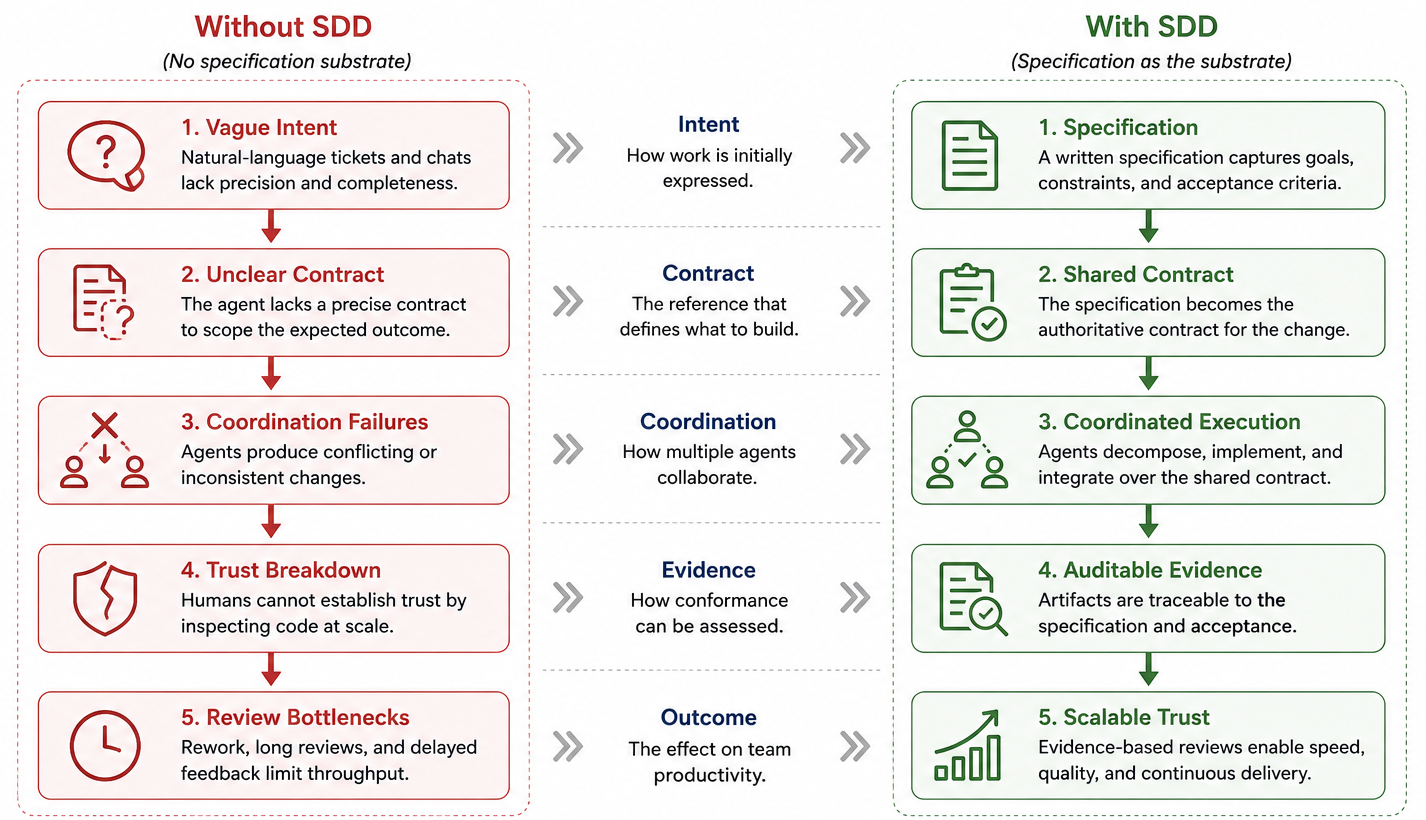}
    \caption{
    Specifications provide the substrate through which contracts, coordination mechanisms, and trust can emerge in agentic software engineering. Without specifications, coordination and trust must be reconstructed from code and conversational history; with SDD, specifications become the shared contract that enables auditable evidence and scalable trust.
    }
    \label{fig:sdd-trust-chain}
\end{figure}

As illustrated in Figure~\ref{fig:sdd-trust-chain}, the deepest reason why SDD enables ASE is that it provides a substrate over which all human--agent interaction can be defined. When the team commits to specifications as the authoritative contract, the recurrent interactions become operations on specifications: a human briefs an agent by writing a specification; an agent requests consultation by citing a specification clause it cannot resolve; a human reviews agent output by auditing evidence against the specification; and a team encodes its standing norms as \emph{normative specifications}. Without this substrate, none of these interactions has a referent; with it, they become definable, teachable, and measurable. Sections~\ref{sec:harness} and~\ref{sec:patterns} build on this substrate.

\section{Harnessing Agentic Capability: The Technical and Methodological Harness}\label{sec:harness}

The notion of \emph{harness} has recently consolidated in the ASE community as the name for everything in an agent system except the model itself, i.e., Agent = Model + Harness \citep{langchain2026harness,fowler2026harness}. The harness is the infrastructure that turns a text-generating model into an actor that can do useful engineering work: the orchestration loop, the tools, the context management, the memory, the guardrails, and the verification mechanisms (e.g Claude Code, OpenCode). In this section, we develop the harness concept in two steps. First, authors characterize the technical harness around the agent (Section~\ref{sec:technical-harness}). Second, being this is one major contribution in this section, authors argue that teams require a second, \emph{methodological} harness, whose mechanisms are described together with worked examples (Sections~\ref{sec:methodological-harness}--\ref{sec:harness-whole}). The examples are deliberately small and concrete. It is worth mentioning that they are illustrative and academic constructions of the authors, not reports of specific industrial systems. While several of the individual mechanisms have appeared independently in recent practitioner literature, our contribution is their organization into a coherent methodological harness centered on specifications as the substrate of coordination.

\subsection{The technical harness}\label{sec:technical-harness}

A language model, by itself, takes text as input and produces text as output. It cannot edit files, run tests, query an issue tracker, or remember yesterday's session. The technical harness supplies these capabilities (e.g., OpenCode, ClaudeCode, Codex, among others). Following the practitioner literature \citep{fowler2026harness,langchain2026harness}, its principal components are the following:

\begin{itemize}
\item \textbf{The orchestration loop,} which repeatedly feeds the model with the current state, parses its output to detect tool calls, executes the tools, and returns the results, until the task is judged complete.
\item \textbf{Tools,} i.e., the actions available to the agent: file editing, shell execution, test running, repository queries, and external integrations (e.g., the issue tracker or the CI system, typically via protocols such as MCP).
\item \textbf{Context management,} which decides what enters the model's limited context window: which files, which rules, which prior decisions. Since context is a scarce resource, this component has a determinant effect on output quality.
\item \textbf{Memory and state persistence,} which provide continuity across sessions, e.g., progress files, session summaries, or committed intermediate states.
\item \textbf{Guardrails and permissions,} which constrain what the agent may do without human confirmation (e.g., which shell commands require approval, which directories are writable).
\item \textbf{Verification loops,} which let the agent check its own work, i.e., compile, run the test suite, run linters, and self-correct before presenting results.
\end{itemize}


Two properties of the technical harness deserve emphasis. First, it is \emph{transient}: as models improve, harness complexity tends to decrease, because capabilities that required scaffolding become native to the model \citep{langchain2026harness}. Second, it is increasingly \emph{co-evolved} with the model: frontier models are post-trained with specific harnesses in the loop, so the harness is not a neutral wrapper but part of the agent's effective behavior. Both properties imply that investments in the technical harness depreciate quickly.

\subsection{The methodological harness}\label{sec:methodological-harness}

The technical harness governs one agent in one session. However, it does not, by itself, solve the problems identified in Section~\ref{sec:vibe}: reproducibility, auditability, transferability, review saturation, and coordination are \emph{team-level} properties. For this reason, authors distinguish a second harness, the methodological harness, whose components are practices and artifacts owned by the team rather than software around the model. Its central artifact is the specification (Section~\ref{sec:sdd}), and its mechanisms are described below. Our claim is that the two harnesses are complementary and asymmetric in durability: the technical harness is transient and depreciates as models improve, whereas the methodological harness encodes the team's intent, norms, and accumulated decisions, and therefore \emph{appreciates} over time. This asymmetry has a practical consequence, i.e., teams should invest preferentially in the durable harness and treat tool selection as a secondary concern.

\begin{figure}[t]
    \centering
    \includegraphics[width=\textwidth]{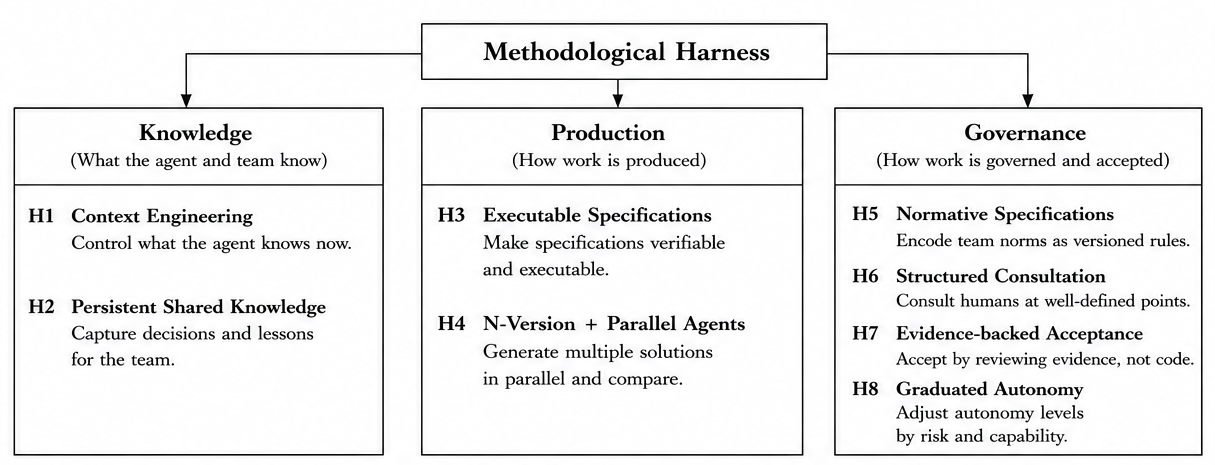}
    \caption{
The methodological harness. The eight mechanisms proposed in this paper are organized into three complementary functions: knowledge management, production support, and governance. Together, they provide the team-level infrastructure required for scalable human--agent collaboration.
}
    \label{fig:mharnes}
\end{figure}

Figure~\ref{fig:mharnes} summarizes the methodological harness and groups its eight mechanisms into three complementary functions: knowledge, production, and governance. The remainder of this section traverses these three groups from left to right, describing eight mechanisms of the methodological harness. For each mechanism, we state the commitment it implies and illustrate it through a worked example. The running example is a feature in a fictitious e-commerce backend: adding a \emph{refund} operation to a payments service. The first mechanism, context engineering, governs what an agent knows within a session; the second one, persistent shared knowledge, governs what the team knows across sessions and agents and is, in our analysis, the key enabler of teamwork in this agentic model.

\subsection{Harness 1: Context engineering}\label{sec:context}

\textbf{Commitment.} The team systematically constructs and maintains the information environment in which agents operate, as a shared and versioned asset rather than as private per-developer configurations.

Context engineering is the team-level counterpart of prompt engineering. The asymmetry between the two is instructive: a poor prompt is detected immediately, because the answer is visibly wrong, whereas poor context is detected weeks later, in production, when the agent has generated plausible code that silently violates conventions or reverts prior decisions. In other words, prompt quality is local and bounded, while context quality is shared and unbounded. 

\textbf{Example.} Suppose the payments service stores all timestamps in UTC as ISO-8601 strings, a decision recorded two years ago in an architecture decision record (ADR) that lives in a wiki the agent never sees. When asked to implement the refund operation, the agent generates code that stores the refund timestamp using the server's local timezone, which is the statistically common pattern in its training data. The code compiles, the tests (which run in UTC in CI) pass, and the defect surfaces months later in reconciliation reports. The context-engineering correction is to move the decision from the wiki into the system specification that the agent loads at session start. A rule such as ``All timestamps are UTC, ISO-8601, suffix \texttt{Z}; never use the system default timezone'' prevents the entire class of defect, for every future feature, at the cost of one line. The general pattern is that each decision documented in agent-readable context converts a recurring review burden into a one-time specification cost.

A second-order effect follows from this mechanism: each feature completed under SDD generates specification artifacts that become context for the next feature, so the context asset compounds. Without shared context, each agent session starts from zero; with it, each session starts from where the team left off.

\subsection{Harness 2: Persistent shared knowledge}\label{sec:persistent}

\textbf{Commitment.} The team maintains a persistent store of decisions, rationale, session summaries, and discoveries that is read at the start of every agent session and written at its end, so that knowledge survives across sessions, across agents, and across the humans on the team.

Context engineering (Section~\ref{sec:context}) addresses what an agent knows \emph{within} a session. However, it does not address the more fundamental problem of the agentic model: agents have no operational continuity \emph{between} sessions. A language model is, by construction, stateless, i.e., each new session begins with no memory of what happened before. If nothing intervenes, every session must rediscover the project's decisions, conventions, and prior dead-ends, which is expensive in tokens, in time, and in the consistency of the result. This is the same discontinuity that vibe coding suffers (Section~\ref{sec:vibe}), now made structural: not only is the reasoning of a past session lost, but the very state of the work is lost when the session closes or the context window is compacted.

Authors argue that solving this discontinuity is the precondition for \emph{teamwork} in the agentic model, and that this is the deepest reason why a methodological harness is necessary. The argument runs as follows. A team is, operationally, a set of agents (human and artificial) that must act on a shared understanding they did not each individually construct. In the pre-AI team, that shared understanding lived in code, documentation, and the memory of colleagues, and it was transferred through review and conversation. In a model where much of the work is performed by stateless agents in isolated sessions, none of those transfer mechanisms operates by default: a sub-agent spawned to implement a task knows nothing of the decision another agent made an hour earlier, and a developer returning the next morning faces an agent that has forgotten the entire prior conversation.

This role resembles the concept of \emph{transactive memory} in organizational theory, whereby groups maintain shared knowledge through distributed but persistent memory structures rather than through the memory of any single individual \citep{wegner1987transactive,ren2008transactive}.

Persistent shared knowledge is the mechanism that reconstitutes this shared understanding. In this way, a collection of stateless sessions can behave as a team rather than as a sequence of unrelated strangers.

Three properties distinguish a persistent-knowledge store that supports teamwork from a mere log. First, it stores \emph{summaries and decisions}, not raw transcripts: the question ``why is the refund idempotency keyed on the request header rather than the body?'' must be answerable in one retrieval, not by replaying a conversation. Second, it supports \emph{progressive (lazy) retrieval}: the agent does not load the entire history, which would itself be context pollution (Section~\ref{sec:context}), but retrieves only what is relevant to the current task, typically through a lightweight ranked search over the store. Third, it is \emph{agent-neutral and tool-neutral}: the same store is read and written by whichever agent or runtime is in use, so that knowledge produced by one tool is available to another. The third property is what makes the store a team asset rather than a per-tool silo; without it, each agent maintains a private memory and the team fragments exactly as it does under vibe coding.

\textbf{Example.} The refund feature is implemented across two days. On the first day, an agent and an engineer establish, after a recorded consultation (Section~\ref{sec:consultation}), that idempotency will be keyed on a client-supplied header and enforced by a unique database constraint; the decision and its rationale are written to the persistent store as the session ends. The next morning, the engineer opens a fresh session, possibly with a different agent or even a different tool, and asks ``where did we leave the refund work?''. The agent retrieves the relevant entries from the store, i.e., the idempotency decision, the open task list, and the note that the listing endpoint still lacks pagination, and resumes without the engineer re-explaining anything. Later that day, when the orchestrator spawns a sub-agent to implement the refund-listing endpoint, that sub-agent retrieves the same idempotency decision from the store and does not contradict it. Hence, the store has carried one decision across a session boundary, across a tool boundary, and across an agent boundary, which is, concretely, what it means for the work to belong to a team rather than to a session.

This mechanism also reframes a cost that the productivity literature underestimates. A substantial share of agentic effort is spent not on code, but on the production and re-production of context: re-explaining the project, regenerating design documents, re-deriving decisions that were made before. A persistent-knowledge store amortizes this cost across sessions. This is both an economic and a quality argument: economic because re-derivation consumes tokens and time, and quality because re-derivation is where silent inconsistency enters. In this sense, persistent shared knowledge is not merely a memory mechanism; it is the organizational substrate that allows a collection of stateless agent executions to behave as a coherent team.

\subsection{Harness 3: Executable specifications}\label{sec:executable}

\textbf{Commitment.} Specifications are written to be consumed by agents and validated by automated checks, not only read by humans.

A specification under SDD is not narrative. The discipline of executability has three operational consequences: acceptance criteria are testable (Given--When--Then is the canonical form), an option that \citet{zhang2025} have already validated by using Gherkin scenarios as a semantic bridge between user requirements and agent-generated code; constraints are checkable (a constraint that no tool can enforce is a wish rather than a constraint); and decisions are explicit, so that conformance checks can validate them downstream.

\textbf{Example.} The refund feature specification includes the following acceptance criterion: \emph{Given} a captured payment of amount $X$, \emph{when} a refund of amount $Y \le X$ is requested with an idempotency key, \emph{then} the refund is executed exactly once, regardless of how many times the request is retried with the same key. This criterion is directly translatable into a test that issues the same request twice and asserts a single ledger entry. The specification also includes the constraint ``must not modify the \texttt{ledger} module'', which a simple ownership check in CI can enforce, and the decision ``expose the operation as \texttt{POST /payments/\{id\}/refunds} returning 201, not as \texttt{PUT}'', which an API-conformance check validates against the OpenAPI description. In our experience, a specification that is consumed by an agent and validated by tests stays alive, whereas one that is only read by humans decays into the kind of documentation nobody trusts.

\subsection{Harness 4: The N-version mindset and parallel agents}\label{sec:nversion}

\textbf{Commitment.} The team treats cheap parallel generation of candidate solutions as a normal mode of working, and develops the evaluative skills it requires.

A characteristic capability of the agentic model is that a single specification can be dispatched to several agents (different models, different decomposition strategies) which return several candidate implementations in parallel \citep{hassan2025agentic}. This changes the developer's cognitive task from ``write the implementation'' to ``specify the contract clearly enough that several acceptable implementations can be compared, and select or compose the best one''. Hence, the required skills are evaluative and combinatorial rather than productive, such as comparative reading of implementations against the same specification, identification of the design trade-offs each one embodies, and recombination of components across candidates.

\textbf{Example.} The refund specification is dispatched to three agents. Candidate A implements idempotency by storing the idempotency key in the refunds table with a unique constraint, relying on the database for atomicity; candidate B uses a distributed lock in Redis; and candidate C checks for an existing refund before inserting, without a constraint. A reviewer who reads the three candidates against the specification can discard C immediately (the check-then-insert pattern has a race condition that violates the exactly-once criterion), and can articulate the trade-off between A (simpler, but it couples idempotency to the relational schema) and B (it adds an infrastructure dependency, but it works across shards). Note that this comparison is only possible because the exactly-once criterion was explicit in the specification; against a vague ticket (``add refunds''), all three candidates look equally plausible.

\paragraph{Working trees, not shared branches} Parallel agent generation raises an isolation problem that is easy to overlook and expensive to get wrong. If several agents operate in the same working directory, they collide: one overwrites a file another is editing, tests fail for reasons unrelated to the feature under work, and the shared version-control index suffers lock contention. Worse, the failures are silent, since an agent does not notice that another has changed the ground under it. The conventional branch model does not solve this problem, because a single working directory can have only one branch checked out at a time; running several agents then forces the team either to serialize them (which defeats the purpose) or to clone the repository several times (which wastes space and breaks the sharing of history). The mechanism that resolves this, and which has become the cornerstone for parallel agentic work during 2026, is the \emph{git working tree} (worktree): each agent is given its own linked working directory, with its own private index and its own branch, while all working trees share a single underlying object store. In this way, the agents are completely isolated at the file level, so an agent's uncommitted edits are invisible to the others and index locks no longer contend; at the same time, they share history and remotes, so creating a working tree is near-instantaneous and consumes negligible disk. Crucially, conflicts move from \emph{active-work time}, where they are silent and corrupting, to \emph{merge time}, where standard tooling detects and surfaces them.

The conceptual point for SDD is that the working tree is the \emph{physical} counterpart of the specification's \emph{logical} isolation: the specification scopes what an agent must produce, and the working tree scopes where it may produce it. Together, they make the N-version regime safe. Thus, three agents implementing the three idempotency candidates above each operate in their own working tree on their own branch, and the team selects or composes from the resulting diffs at merge time, having never risked one agent corrupting another's work. We note, however, that working-tree isolation is isolation of \emph{code}, not of \emph{runtime}: parallel agents still share ports, databases, and external services unless those are also separated (per-tree environment files, scratch databases, port ranges), which is a practical caveat that teams discover quickly.

This mechanism connects directly to delivery discipline, including continuous delivery practices that are central to DevOps culture, as we previously described in \citep{deliverydiaz}. Since each parallel unit of work already lives on its own branch, the team can deliver large changes as a sequence of small, reviewable units rather than as a single oversized pull request. A change that would otherwise arrive as a thousand-line diff, which is a review debt paid by a human colleague rather than a productivity gain, can be partitioned into chained branches, each of them a coherent step that references the same specification. The choice of delivery strategy, i.e., whether to stack units directly onto the mainline (main branch) or onto an intermediate integration branch from which the whole feature can be rolled back atomically, is itself a decision that the harness should make explicit rather than leave to an individual's improvisation.

\subsection{Harness 5: Normative specifications}\label{sec:normative}

\textbf{Commitment.} The team's standing norms, i.e., the conventions, patterns, and prohibitions that every change must respect, are written as an explicit, versioned, reviewed part of the system specification, rather than left tacit in senior engineers' heads.

In our framing, this mechanism is not a separate artifact but a \emph{kind of specification}. The two-level discipline of Section~\ref{sec:definition} distinguished feature specifications (what is being built now) from system specifications (the durable world the project lives in). Normative specifications are the normative content of the latter: while most of a system specification describes how the system \emph{is} (its architecture and models), the normative part prescribes how every change \emph{must behave} (its conventions and constraints). Since they live in the always-loaded system specification, materialized in the rule file of Section~\ref{sec:definition}, they govern agent behavior across every session without anyone invoking them. This is precisely what distinguishes them from skills (Section~\ref{sec:definition}): a normative specification is declarative and always active (``what to respect''), whereas a skill is procedural and loaded on demand (``how to do a particular thing''). A team that places a standing convention in an on-demand skill will find the agent violating it in every session that does not happen to trigger the skill. This is why normative specifications are a natural consequence of SDD rather than an independent practice.

Framing these norms as a kind of specification, rather than as a free-standing artifact, has the advantage of inheriting the discipline the rest of the harness already imposes: norms are added by reviewed pull request, deprecated when they no longer hold, and checked for compliance. It is the same practice that \citet{hassan2025agentic} describe under the name \emph{mentorship-as-code}. We adopt their underlying insight, i.e., that institutional memory must become explicit machine-readable guidance, but we situate it inside our specification substrate rather than treating it as a distinct construct, which is what makes the rule file, the system specification, and the team's norms one and the same artifact viewed from three angles (Section~\ref{sec:harness-whole}).

\textbf{Example.} During review of agent output, a senior engineer repeatedly applies the same correction: agent-generated service methods catch generic exceptions and log them, whereas the project's convention is to map low-level exceptions to typed domain errors at the service boundary. After the third repetition, the correction is generalized into a normative clause and added, by reviewed pull request, to the system-specification rule file (\texttt{agent.md} or equivalent): ``Service-layer methods must not catch \texttt{Exception}; catch specific exceptions and map them to subclasses of \texttt{DomainError}, as in \texttt{PaymentService.capture()}.'' The clause is paired with a lint sensor (Section~\ref{sec:technical-harness}) whose violation message repeats the rule text, so the agent self-corrects before a human ever sees the code. Note that this clause belongs in the always-loaded rule file, not in an on-demand skill, since it must hold in \emph{every} session that touches the service layer, including the ones that never invoke a skill. A practical corollary is that such clauses can be proposed by the agent itself: when it observes that a human repeatedly refactors its output along the same pattern, it can draft the generalized clause for the team to approve, which surfaces tacit norms the team had never written down.

\subsection{Harness 6: Structured consultation}\label{sec:consultation}

\textbf{Commitment.} Human involvement during agent execution is structured around scoped, recorded consultation rather than continuous supervision.

Humans cannot keep pace with parallel agent throughput, and forcing continuous supervision recreates the bottleneck that the agentic model was meant to relieve. Instead, the agent suspends execution when it encounters a decision it cannot or should not resolve, and raises a scoped consultation request that includes the question, the options considered, the relevant specification clauses, and the consequences of each option. This mechanism adopts the Consultation Request Pack that \citet{hassan2025agentic} propose as a first-class artifact of ASE; our contribution is to ground it in the specification substrate, i.e., the request cites the specification clause it cannot resolve, and the human's resolution is recorded against that specification, so that the same question is not asked twice and the decision becomes part of the contract itself. 

\textbf{Example.} While implementing the refund listing endpoint, the agent finds that the specification requires pagination but does not state the strategy, and the codebase contains both offset-based and cursor-based precedents. Instead of improvising, it raises a consultation: ``Specification \#4.2 requires pagination but does not fix the strategy. Option 1: offset-based, consistent with the older \texttt{orders} endpoints. Option 2: cursor-based, consistent with the newer \texttt{payments} endpoints and stable under concurrent inserts. Refunds are append-mostly, which favors Option 2.'' The responsible engineer answers in one line, the resolution is committed alongside the specification, and the next agent that touches pagination finds the decision already made. Note that consultation differs from review in its position in time: review reads after the fact, whereas consultation decides before the fact, on a question the agent has surfaced. In this way, human attention is concentrated at the points of highest leverage.

\subsection{Harness 7: Evidence-backed acceptance}\label{sec:evidence}

\textbf{Commitment.} The unit of completion is not ``the PR is merged'' but a structured evidence bundle demonstrating conformance to the specification.

In Agile, done means that the PR is merged after a human read the code. In vibe coding, done means that it feels right. Neither of them scales to agent volumes. Under SDD, the deliverable is an evidence pack that demonstrates conformance along the five merge-readiness criteria that \citet{hassan2025agentic} define for their Merge-Readiness Packs: functional completeness (acceptance criteria exercised), verification soundness (the test plan is coherent, not merely green), engineering maintainability and clean code (code smells, style and pattern conformance), rationale (a human-readable summary of the approach and trade-offs), and auditability (links to the originating specification, the rules in force, and the agent trajectory and decision). What SDD adds to their proposal is the referent: under our framing, every item of evidence discharges a clause of the originating specification, which is what makes the audit tractable.

\textbf{Example.} The refund feature arrives for review as an evidence pack containing: the executed acceptance tests, including the double-submission idempotency test, with their mapping to specification clauses; the CI (continuous integration) output of automated checks, including validation of ownership boundaries in the ledger and conformance to the API specification; a three-paragraph rationale explaining the choice of the unique-constraint idempotency design over the lock-based alternative, with a pointer to the recorded consultation; and links to the specification version and the normative specifications in force at generation time. The reviewer's task changes from reading several hundred lines of unfamiliar code to auditing whether the evidence actually discharges the contract, drilling into specific artifacts only where the evidence is weak. Per artifact, this is less cognitively demanding and more informative per minute spent; at the team level, it is what makes the review bottleneck of Section~\ref{sec:introduction} structurally addressable.

\subsection{Harness 8: Graduated autonomy}\label{sec:autonomy}

\textbf{Commitment.} Autonomy is calibrated per class of task, explicitly and revisably, rather than granted or denied globally.

An agent given full autonomy will sometimes deliver well and sometimes drift; an agent given none is a slow autocomplete. \citet{hassan2025agentic} describe autonomy as a ladder rather than a switch. At team scale, what matters is that the autonomy level for each class of work is an explicit, versioned decision.

\textbf{Example.} The team's workflow definitions state that dependency patch updates run fully autonomously (the sensors, i.e., the test suite and the vulnerability scanner, are judged sufficient); that ordinary feature work like the refund endpoint runs autonomously with mandatory consultation on unresolved specification ambiguities and evidence-pack review at the end; and that any change touching the authentication module requires a human-approved plan before execution and step-wise review. When, months later, the team observes that dependency updates have run for a quarter without incident while two authentication changes raised serious consultation questions, the calibration is revised in one direction and confirmed in the other. The point is that these are team decisions recorded in artifacts, not individual habits.

\subsection{The harness as a whole}\label{sec:harness-whole}

\begin{figure}[t]
    \centering
    \includegraphics[width=\textwidth]{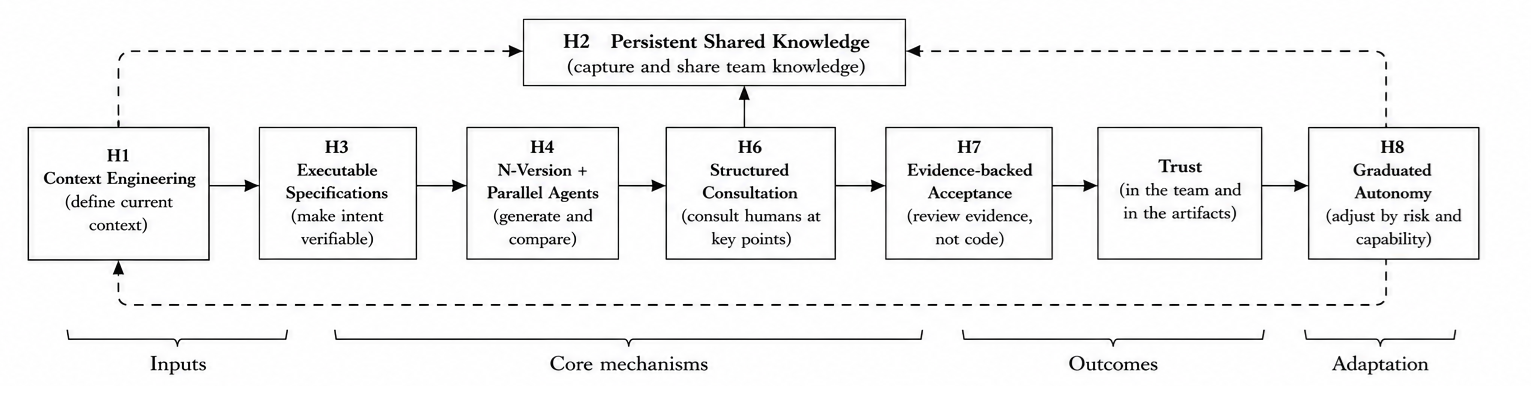}
    \caption{
The methodological harness as an integrated system. Persistent shared knowledge (H2) closes the loop by capturing decisions, rationale, and outcomes, allowing subsequent agents and sessions to build upon previous work rather than restarting from scratch.
}
    \label{fig:pkharnes}
\end{figure}

Figure~\ref{fig:pkharnes} illustrates how the individual mechanisms interact as a coherent socio-technical system rather than as independent practices.

As illustrated in Figure~\ref{fig:pkharnes}, the eight mechanisms described above are not independent additions to a toolkit; rather, they reinforce each other. Executable specifications make evidence-backed acceptance possible; evidence-backed acceptance absorbs the volume that N-version generation produces; N-version generation, made safe by working-tree isolation, requires explicit autonomy calibration; calibration is stabilized by normative specifications; these norms accumulate through review and consultation; the accumulation is written to the persistent-knowledge store; and the store, together with the engineered context, is what lets the next session, the next agent, and the next teammate begin where the last one left off. Persistent shared knowledge is, in this sense, the mechanism that closes the loop: without it, the other mechanisms still operate within a session, but their products do not survive to compound across the team. Hence, a team that adopts the harness in fragments, e.g., writing feature specifications but neither encoding its norms nor persisting its decisions, obtains a fraction of the benefit while paying most of the discipline cost. This observation is consistent with the empirical pattern that partial, ungoverned adoption can degrade team outcomes \citep{dora2024}, and it implies that adoption should be studied as a staged organizational process; we return to this point in the research agenda (Section~\ref{sec:agenda}).

A consequence of this integration deserves to be stated explicitly, since it collapses three notions that the practitioner discourse keeps separate. The rule file that the team maintains (e.g., \texttt{agent.md}) is, simultaneously, (i) the \emph{materialization of the system specification} (Section~\ref{sec:definition}), (ii) a \emph{guide} in the feedforward sense of the technical harness (Section~\ref{sec:technical-harness}), and (iii) the carrier of the team's \emph{normative specifications} (Section~\ref{sec:normative}). In other words, these are not three artifacts but one artifact fulfilling three complementary roles: a specification when we ask what the durable world is, a guide when we ask how it steers the agent before it acts, and a body of norms when we ask what every change must respect. This unification is, in our view, one of the conceptual simplifications that SDD brings to a discourse that currently treats rule files, agent guidance, and conventions as unrelated artifacts.

\section{Human--Agent Interaction Patterns and the Redefinition of Human Roles}\label{sec:patterns}

The harness has a corollary: the team itself transforms. This section identifies five recurring human--agent interaction patterns and then describes how the human role is redefined around them, principally as a shift from authoring code to orchestrating, specifying, and verifying the work of agents.

\subsection{Five interaction patterns}\label{sec:five-patterns}

The five patterns were derived from the qualitative synthesis described in Section~\ref{sec:methodology} and summarized in Table~\ref{tab:evidence-traceability}. They are not intended to be exhaustive; rather, they are the recurring interaction categories whose mastery, in our analysis, distinguishes mature practice. Note that each pattern is an operation on the specification substrate of Section~\ref{sec:substrate}.

\begin{figure}[!htbp]
    \centering
    \includegraphics[width=\textwidth]{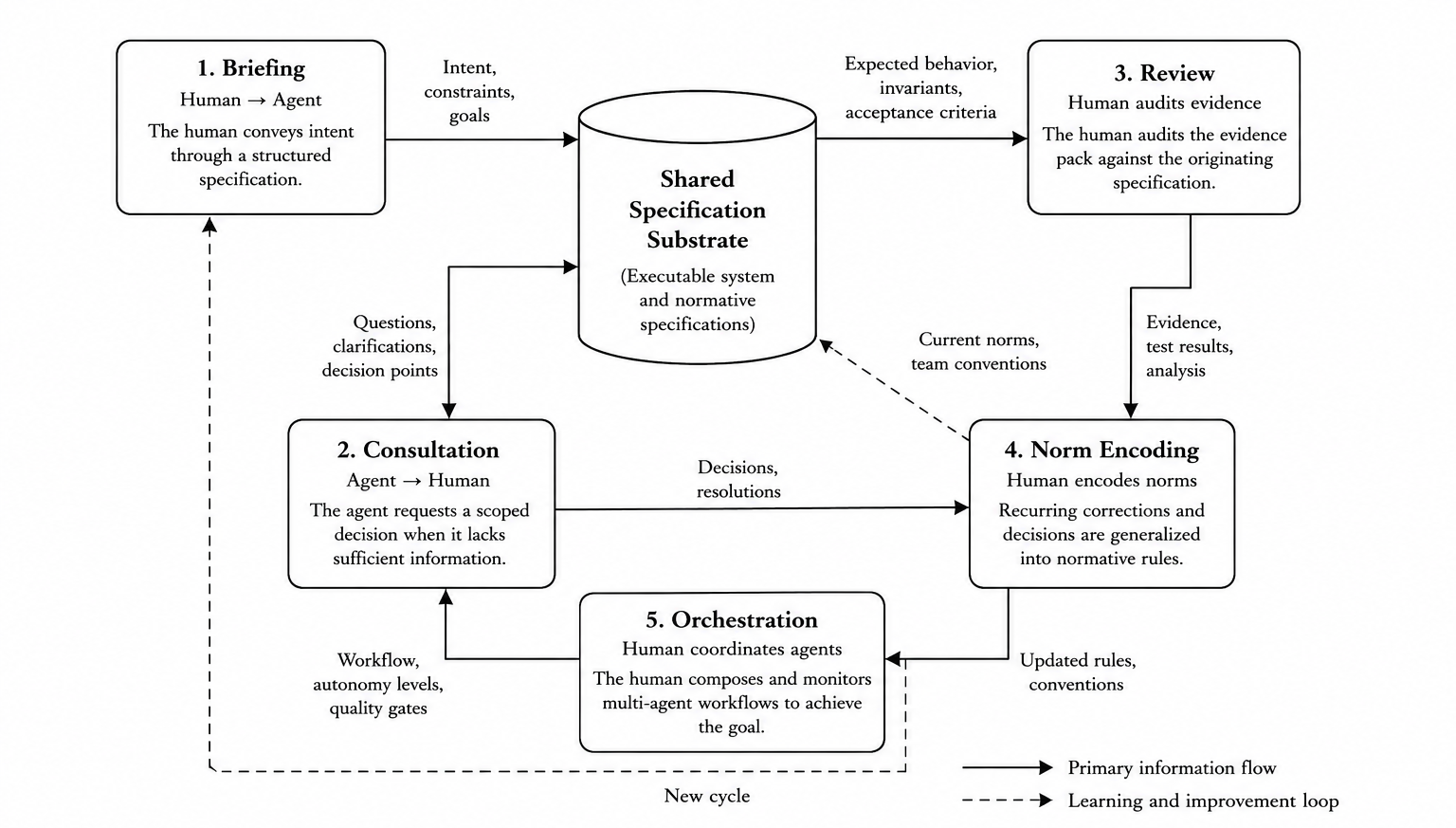}
    \caption{
    The five recurring human--agent interaction patterns under SDD. Each pattern operates on the shared specification substrate, creating a continuous improvement cycle in which specifications are created, interpreted, validated, enriched, and reused across successive agent executions.
    }
    \label{fig:interaction-patterns}
\end{figure}

Figure~\ref{fig:interaction-patterns} summarizes the five recurring interaction patterns. Rather than isolated activities, they form a continuous collaboration cycle centered on the shared specification substrate.

\paragraph{Pattern 1: Briefing} The human conveys intent through a structured specification that the agent then operates under. The interaction is upstream, human-authored, and contractual. It demands of the human the ability to articulate intent precisely, including constraints and edge cases, and to think in properties (``the operation must be idempotent'') rather than in examples. It demands of the team a shared template for briefs \citep{hassan2025agentic} and a review practice for the briefs themselves, not only for the resulting code.

\paragraph{Pattern 2: Consultation} The agent suspends and requests a scoped decision (Section~\ref{sec:consultation}). The interaction is agent-initiated, bounded, and recorded. It demands of the human rapid context loading and the willingness to contribute expertise when invoked, rather than remaining in a constant supervisory role. It demands of the team routing rules (i.e., which roles handle which classes of consultation) and response-time discipline, since an unanswered consultation blocks parallel agents.

\paragraph{Pattern 3: Review} The human audits an evidence pack against the originating specification (Section~\ref{sec:evidence}). It demands of the human the ability to read evidence rather than re-derive correctness, and the discrimination to drill selectively into weak evidence. It demands of the team quality gates that produce the evidence automatically, as well as the cultural acceptance of evidence-based review, which some senior engineers initially experience as a loss of control.

\paragraph{Pattern 4: Norm encoding} The human turns recurring corrections into normative specifications in the system-specification rule file (Section~\ref{sec:normative}). The interaction is cumulative and team-shared. It demands of the human the ability to generalize from specific corrections to standing clauses, and it demands of the team an authoring and maintenance discipline for the normative content of the system specification.

\paragraph{Pattern 5: Orchestration} The human composes a multi-agent workflow for a goal: which agents, in what order, with what autonomy levels and evidence requirements. A useful distinction here, prominent in practitioner toolchains, is the one between the \emph{orchestrator} and the agents it coordinates: the orchestrator decides the flow, controls the quality gates, synthesizes results, and keeps final responsibility, but it does not itself write the code, which is produced by the appropriate sub-agent in its own isolated context (Section~\ref{sec:persistent}, Section~\ref{sec:nversion}). This separation keeps the orchestrator's own context uncontaminated by execution detail, which matters because an orchestrator that has to reconstruct the whole history at every step becomes itself a bottleneck. Orchestration demands portfolio thinking about agent capabilities, cost awareness, and the ability to debug workflows rather than code; it demands of the team an orchestration substrate and observability into multi-agent runs.

\subsection{From coder to orchestrator: how the human role changes}\label{sec:roles}

The five patterns describe interactions; taken together, they describe a change in what the human's primary work \emph{is}. In the GenAI-augmented model, the human's primary output was code, with the assistant accelerating its production. Under SDD-governed ASE, the human's primary outputs are briefs, rules, workflow definitions, consultation resolutions, and evidence judgments, i.e., the artifacts of the specification substrate, while code production is largely delegated. Hence, the cognitive center of the work moves upstream (clarity of intent, architectural judgment) and sideways (comparative evaluation of candidate implementations), and away from the line-by-line authoring that previously defined seniority. It is worth noting that this displacement is not an idiosyncratic claim of ours. Among the projections for 2030 that close their roadmap, \citet{amalfitano2026} anticipate a redefinition of the ``10x developer'' as the engineer who effectively coordinates a fleet of agents rather than the one who writes more code, and they expect performance to be measured in agents coordinated rather than in lines produced. What this paper adds to that projection is its operational content, i.e., which artifacts the orchestrator authors, which competences the shift demands, and which team-level mechanisms make it sustainable.
The key change is not that humans stop programming altogether, but that their primary value progressively shifts from producing code to governing the conditions under which agents produce reliable software.










Authors deliberately refrain from coining a new job title for the human who masters this shift; the practitioner and academic communities have proposed several, and the vocabulary is neither stable nor, in our view, the substantive contribution. What matters is the \emph{set of competences} that the shift requires, which differs qualitatively from those of the GenAI-augmented developer:

\begin{itemize}
\item \textbf{Specification literacy:} writing structured text that survives interpretation by both humans and agents, a skill close to technical writing but disciplined toward executability (Section~\ref{sec:executable}).
\item \textbf{Architectural reasoning as a daily competence,} because the human now makes, routinely, the design decisions that an agent cannot be trusted to make.
\item \textbf{Comparative reading:} evaluating several candidate implementations against one specification, a task closer to grading than to coding (Section~\ref{sec:nversion}).
\item \textbf{Norm articulation:} expressing standing conventions clearly enough for them to become normative specifications, not merely enforcing them case by case (Section~\ref{sec:normative}).
\item \textbf{Orchestration and cost awareness:} composing and debugging multi-agent workflows, and accounting for the computational and financial cost they incur (Section~\ref{sec:autonomy}, Section~\ref{sec:five-patterns}).
\item \textbf{Strategic patience:} resisting the temptation to ``just code it'' when the situation calls for specifying it, which is the cultural inversion that the whole agentic model demands.
\end{itemize}

How these competences are distributed across people, whether they consolidate into a single new role or are spread across redefined existing ones, and how engineers acquire them, are open empirical questions that authors list in the research agenda for future work (Section~\ref{sec:agenda}).

The shift toward orchestration is visible across every conventional role. The developer role bifurcates: some developers move toward orchestration and specification work, while others remain hands-on for work that resists agentic execution (novel domains, complex performance optimization, regulatory edge cases). The technical lead becomes the primary author of system specifications, whose impact scales across all features, since each specification is reused rather than re-created. The product manager becomes a co-author of product specifications in a tighter loop than under Agile, because ambiguity at the product level now propagates directly into agent output. The QA role inverts, from after-the-fact testing to before-the-fact specification of testability. Finally, operations and security roles shift from gatekeeping to authoring the operational and security specifications that govern the corresponding agents, according to previous trends on DevSecOps, as we proposed in \citep{devsecopsdiaz}.

\subsection{N-to-N topology and the cultural shifts}\label{sec:topology}

\begin{figure}[!htbp]
    \centering
    \includegraphics[width=\textwidth]{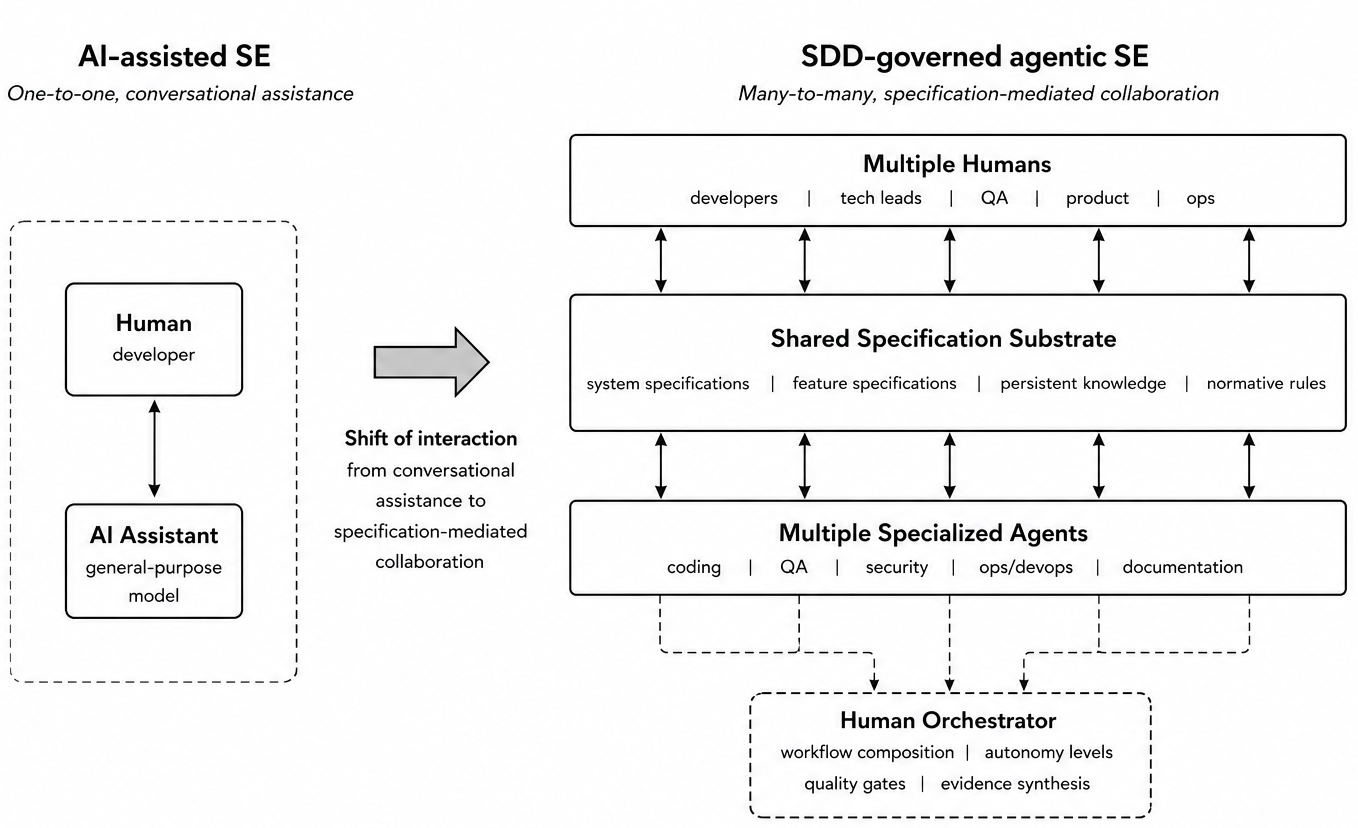}
    \caption{
    Evolution of human--AI collaboration. GenAI-assisted software engineering relies on a one-to-one interaction between a developer and an assistant. Under SDD-governed agentic software engineering, multiple humans and specialized agents collaborate through a shared specification substrate, while orchestration governs workflows, autonomy levels, quality gates, and evidence synthesis.
    }
    \label{fig:n2n-topology}
\end{figure}

The resulting topology is N-to-N, i.e., multiple humans collaborating with multiple agents through the specification substrate, with persistent shared knowledge (Section~\ref{sec:persistent}) as the medium that lets them act as one team rather than as disconnected sessions. This is an evolution of our previous work on DevOps Team Structures \citep{lopez2021}. Figure~\ref{fig:n2n-topology} summarizes this shift from one-to-one conversational assistance to many-to-many, specification-mediated collaboration. Unlike the traditional one-to-one interaction between a developer and an assistant, SDD enables a many-to-many collaboration model in which humans and specialized agents interact through shared specifications and persistent knowledge rather than through isolated conversations. Three operational implications follow. First, cross-human handover: a consultation raised by one agent may be resolved by a different human than the one who authored the brief and specifications, which requires routing rules and expertise mapping. Second, cross-agent handover: an agent may need to invoke a specialist agent for a sub-task, which requires shared specification access and shared knowledge. Third, hybrid composition: specialist functions may be filled by humans, agents, or pairs, so team composition becomes a design variable.

Finally, three cultural shifts deserve explicit mention, since adoption reports repeatedly identify them as harder than the technical work. The first one is the \emph{deference shift}: seniority progressively shifts away from coding excellence toward specification and architectural judgment, a change that challenges established identities and is a common source of resistance among senior programmers. The second one is the \emph{accountability shift}: under SDD, the brief's author is accountable for the specification, the orchestrator for the workflow, and the reviewer for the evidence audit; organizations whose performance management assumes individual code ownership must therefore reconfigure. \citet{amalfitano2026} document this same displacement, which they characterize as a reversal of code ownership induced by GenAI, and they place intellectual property, licensing, and legal liability among the cross-cutting issues that remain open across every form of augmentation. Hence, the reconfiguration we describe is not merely internal to the team, since it also has contractual and regulatory dimensions that we do not address here. The third one is the \emph{skill investment shift}: junior engineers can no longer become senior only by writing code, so career frameworks must incorporate specification authoring, workflow design, and the encoding of team norms, which connects to the calls to re-imagine SE education \citep{hoda2026agentic}.

\section{Traceability of the Synthesis}\label{sec:traceability}

Having introduced the specification substrate, the harness mechanisms, and the interaction patterns, Table~\ref{tab:evidence-traceability} makes explicit how the synthesis links source evidence to the constructs proposed in this paper. The table is representative rather than exhaustive; the full extraction matrix and source-selection material are provided in the replication package.

\begin{table*}[t]
\centering
\caption{Representative traceability from source evidence to synthesized mechanisms and interaction patterns.}
\label{tab:evidence-traceability}
\scriptsize
\begin{tabularx}{\textwidth}{p{3.0cm}p{3.3cm}p{2.7cm}X}
\toprule
\textbf{Source evidence} & \textbf{Extracted concept} & \textbf{Category} & \textbf{Resulting mechanism/pattern} \\
\midrule
Industrial productivity and code-quality reports \citep{farosai2025paradox,dora2024,dora2025,metr2025,gitclear2025,stackoverflow2025} & Local acceleration can coexist with review latency, instability, defects, and maintainability degradation. & Team-level governance gap & Motivation for SDD as a team-scale discipline and for evidence-backed acceptance (RQ1). \\
\midrule
Harness practitioner accounts \citep{fowler2026harness,langchain2026harness} & Useful agent behavior depends on tools, orchestration, context, memory, guardrails, and verification around the model. & Technical harness & Technical harness around the agent and its relation to methodological controls (RQ1). \\
\midrule
ASE roadmaps and vision papers \citep{hassan2025agentic,hoda2026agentic,amalfitano2026} & Agentic work requires workflow definitions, oversight, merge-readiness evidence, and role redefinition. & Methodological harness & Specification substrate, structured consultation, graduated autonomy, and redefined human roles (RQ1--RQ2). \\
\midrule
Requirement-centered generation and executable scenarios \citep{zhang2025} & Requirements can be operationalized as scenarios that bridge intent, generation, and tests. & Executable specification & Briefing pattern and executable-specification mechanism (RQ1--RQ2). \\
\midrule
Merge-readiness, consultation, and autonomy proposals \citep{hassan2025agentic} & Agentic workflows need recorded consultations, evidence packs, and calibrated autonomy levels. & Recorded decision and evidence control & Consultation, review, evidence-backed acceptance, and graduated autonomy (RQ1--RQ2). \\
\midrule
Agentic-teammate and future-role accounts \citep{li2026teammates,amalfitano2026,hoda2026agentic} & Human value shifts from line-by-line production toward coordinating agents and judging outputs. & Human orchestration & Orchestration pattern and transition from coder to orchestrator (RQ2). \\
\bottomrule
\end{tabularx}
\end{table*}

\section{Discussion}\label{sec:discussion}

\subsection{Expected benefits}\label{sec:benefits}

\begin{figure}[!htbp]
    \centering
    \includegraphics[width=0.8\textwidth]{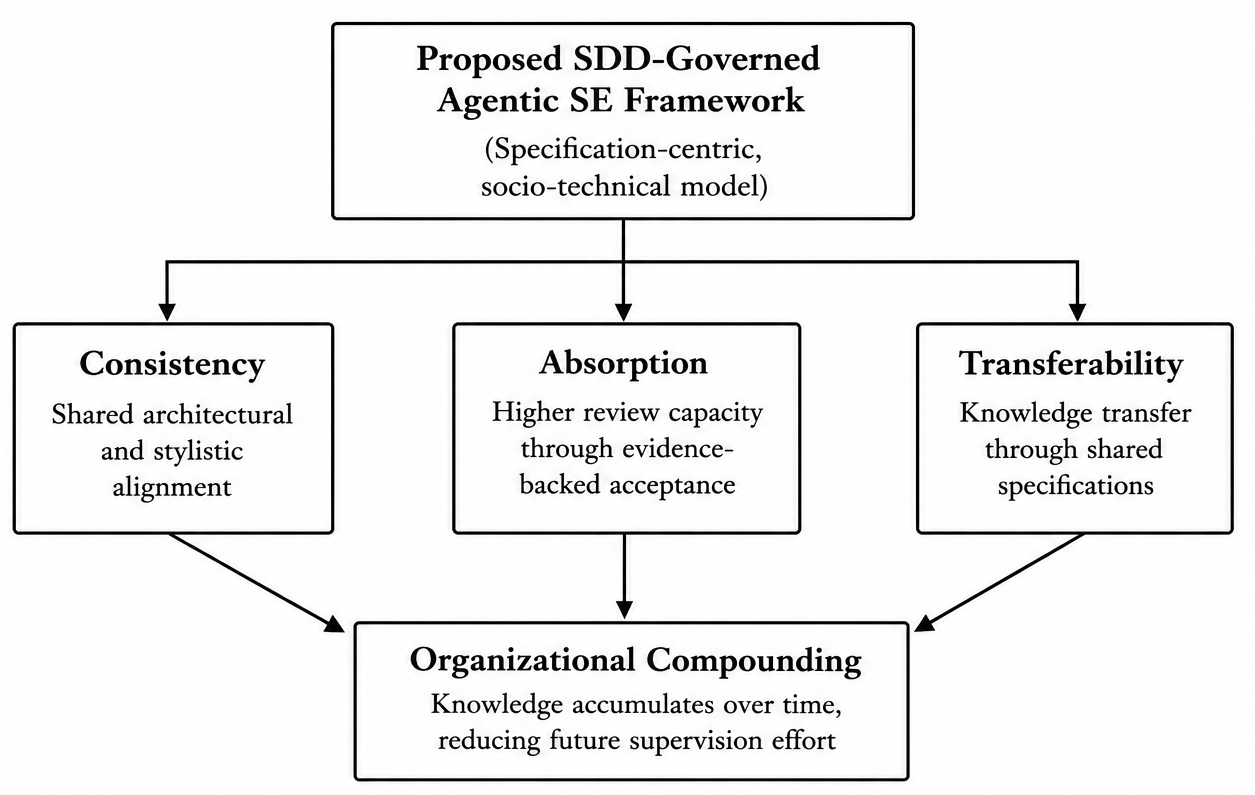}
    \caption{
    Expected organizational benefits of the proposed ASE framework. Consistency, absorption, and transferability improve the current development cycle, while their cumulative effect produces organizational compounding through the continuous growth of shared specifications and persistent knowledge. These benefits are predictions of the proposed framework and remain to be validated empirically.
    }
    \label{fig:expected-benefits}
\end{figure}

This agentic model predicts four benefits, all of which are testable. The first one is consistency: shared context and normative specifications reduce the stylistic and architectural divergence that individual GenAI use produces. The second one is absorption: evidence-backed acceptance addresses the review bottleneck, which the industrial reports identify as the main constraint \citep{farosai2025paradox,qodo2025}. The third one is transferability: onboarding and handover read specifications and the persistent-knowledge store rather than reconstructing intent from code, which is the property that makes the work belong to a team rather than to an individual (Section~\ref{sec:persistent}). The fourth one is compounding: the persistent context asset grows with each feature, so the marginal cost of agent supervision should decrease over time. Authors emphasize that these are predictions of the framework, not yet validated findings. Figure~\ref{fig:expected-benefits} summarizes the four expected organizational benefits predicted by the proposed framework. 

Authors emphasize that these benefits are theoretical predictions derived from the proposed socio-technical framework rather than empirically validated findings. Confirming or refuting these predictions requires empirical evidence from organizations adopting SDD-governed ASE under real development conditions.

\subsection{Expected risks}\label{sec:risks}

The proposed framework also introduces organizational and technical risks that should be considered during adoption. Unlike the expected benefits, these risks arise primarily from changes in governance, organizational practices, and the redistribution of responsibilities between humans and agents rather than from limitations of the GenAI models themselves. Table~\ref{tab:risks} summarizes the principal risks identified together with their corresponding mitigation strategies. 

\begin{table}[t]
\caption{Main risks associated with the transition to SDD-governed ASE.}
\label{tab:risks}
\footnotesize
\begin{tabular*}{\textwidth}{@{\extracolsep\fill}p{2.5cm}p{4.2cm}p{3.1cm}@{}}
\toprule
\textbf{Risk} &
\textbf{Description} &
\textbf{Potential mitigation} \\
\midrule

Talent bifurcation &
Uneven acquisition of orchestration and specification competences across engineers &
Training, mentoring, and gradual reskilling \\

Cultural fracture &
Resistance to changes in roles, seniority, and established development practices &
Incremental adoption and organizational change management \\

Cost spiral &
Uncontrolled growth in computational and engineering costs associated with agent execution &
Cost-aware orchestration, monitoring, and execution policies \\

Specification drift &
Progressive divergence between specifications and generated artifacts &
Continuous synchronization, traceability, and automated governance \\

Platform lock-in &
Dependence on proprietary agent ecosystems or specification formats &
Platform-neutral specifications and version-controlled artifacts \\

\bottomrule
\end{tabular*}
\end{table}

\paragraph{Talent bifurcation} A competency gap may open between the engineers who acquire the orchestration and specification competences (Section~\ref{sec:roles}) and those who do not, with effects on compensation and internal equity; the mitigation requires deliberate training rather than reliance on individual initiative. 

\paragraph{Cultural fracture} Resistance from experienced engineers, driven by the deference shift, can delay adoption even when the technology is technically mature. Similar organizational resistance has been widely observed in previous software engineering transformations, including the transition toward DevOps, where cultural and organizational barriers often proved more difficult to overcome than the supporting technologies themselves \citep{forsgren2018accelerate,diaz2021many}. Successful adoption therefore requires deliberate change management, incremental deployment, and explicit organizational support rather than reliance on individual initiative. 

\paragraph{Cost spiral} Agent invocations consume compute and engineering time; without cost accounting integrated in the orchestration substrate, agentic workflows can become more expensive than the baseline they replace. 

\paragraph{Specification drift} When specifications and code diverge silently, the discipline collapses from within; drift detection and periodic audits are required, and authors see here a natural role for automated governance. 

\paragraph{Platform lock-in} Specifications tightly coupled to a specific agent platform create a new vendor dependence; platform-neutral formats and version control independent of the agent runtime are the mitigation.

This observation reinforces that SDD should be regarded as a governance discipline rather than as a universal prescription. The appropriate degree of specification depends on the uncertainty, stability, and expected lifetime of the work product. SDD asks teams to invest upstream; for exploratory work whose requirements are genuinely unknown, this investment may be premature, and a deliberately lighter discipline (closer to supervised vibe coding, with throwaway artifacts) may be more appropriate. 

\subsection{Threats to validity}\label{sec:threats}

As this work proposes a conceptual socio-technical framework rather than an empirical evaluation, the principal threats concern the interpretation, scope, and evidential basis of the proposed model. The central terms (agentic SE, vibe coding, harness, specification) are recent and controversial. Our definitions stabilize them for this paper, but the community has not converged, and results obtained under our definitions may not transfer to others. 


The agentic model described here is a conceptual synthesis of three independent streams (the academic agentic-SE literature, the specification-centric tradition, and the gray literature and industrial reports). Authors believe that the synthesis is coherent, but it has not yet been validated empirically as a unified model. In particular, whether SDD adoption can mitigate the productivity paradox remains an empirical question that requires further validation. Furthermore, 
agent capabilities are improving rapidly, and parts of the operational detail (artifact schemas, autonomy thresholds, the thickness of the technical harness) will date quickly. Authors argue that the agentic model's core commitments, i.e., specification as contract, evidence-backed acceptance, and the human as orchestrator and verifier, are stable under model improvement, precisely because they concern the team rather than the model. Future empirical work should therefore distinguish between implementation details that may evolve with advances in foundation models and the socio-technical principles proposed here, which are intended to remain stable across technological generations.

Finally, authors discuss these limitations explicitly to delimit our claims based on some quality criteria defined by \cite{lincoln:1985} for qualitative research:

\paragraph{Credibility}
 It is also referred to as trustworthiness, i.e., the extent to which conclusions are supported by rich, multivocal evidence. The strategy to mitigate this threat was to define quality criteria to document inclusion in the research and consensus among researchers in each iteration. Also, we considered both academic and gray literature. The selection of documents was iterative and can be considered a combination of: (i) “Convenience sampling” as we were restricted to the little literature on the phenomenon to be studied. (ii) “Theoretical sampling”, in the sense that we chose which data to collect based on the concepts and categories relevant (agentic SE, vibe coding, harness, specification). (iii) Finally, “Maximum variation sampling”, in the sense that we tried to choose highly diverse documents in our sample. We have triangulated sources and methods. The proposed harness mechanisms are not arbitrary; rather, they arise from the convergence of industry reports, academic roadmap proposals, and observations of current agent-based tools.

\paragraph{Resonance}
It is the extent to which the study’s conclusions make sense to (i.e., resonate with) the community. The three contributions, i.e., the socio-technical model of SDD, the operational characterization of the harness with worked examples, and the typology of interaction patterns through which the human role is redefined, are in line with current needs in the field of agent engineering.

\paragraph{Transferability} 
It shows whether the findings could plausibly apply to other situations. The motivating evidence skews toward web, SaaS, and platform engineering. Generalization to embedded, safety-critical, and scientific software is plausible but unproven; in regulated domains, the evidence-pack mechanism may need to satisfy externally imposed assurance standards that authors have not analyzed.

\paragraph{Usefulness} 
It refers to the extent to which a study provides actionable recommendations to researchers, practitioners, or educators and the degree to which results extend our cumulative knowledge. The agentic model described here is a conceptual synthesis of three independent streams (the academic agentic-SE literature, the specification-centric tradition, and the gray literature and industrial reports). The claim that the adoption of SDD resolves the productivity paradox is still a hypothesis; however, if confirmed, it would provide the industry with a tool to eliminate a bottleneck which, at present, has a negative impact on delivery speed and competitiveness.

\paragraph{Dependability}
It implies that the research process is systematic, well documented, and can be traced. The implementation of the research process has been described using a MLR that any researcher can replicate.

\paragraph{Confirmability}
It assesses whether the findings emerge from the data collected from cases and not from preconceptions. Sections~\ref{sec:sdd}--\ref{sec:traceability}, which address RQ1 and RQ2, describe the chains of evidence that support these results and allow any researcher to verify them.



\section{Research Agenda}\label{sec:agenda}

The agentic model described here identifies five lines of research that, in our view, deserve community attention.

\begin{figure}[!htbp]
    \centering
    \includegraphics[width=\textwidth]{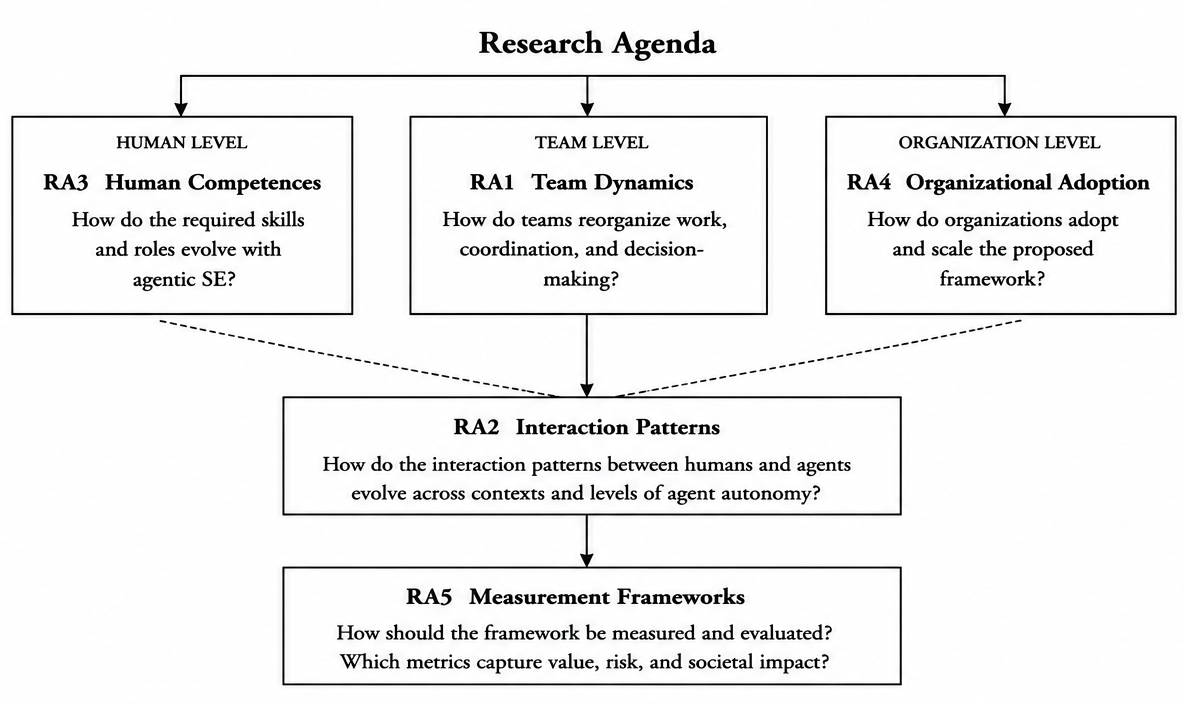}
    \caption{
    Research agenda derived from the proposed socio-technical framework. The agenda spans three complementary levels of analysis: human competences (RA3), team dynamics (RA1), and organizational adoption (RA4). These are connected through the study of human--agent interaction patterns (RA2), while measurement frameworks (RA5) provide the empirical basis for evaluating the effectiveness, risks, and long-term impact of SDD-governed ASE.
    }
    \label{fig:research-agenda}
\end{figure}

Figure~\ref{fig:research-agenda} summarizes these research directions according to the different levels of analysis implied by the proposed framework, from individual competences to organizational adoption and measurement.

\paragraph{RA1. Team dynamics under agentic SE} How do teams reorganize around the orchestration role described in Section~\ref{sec:roles}? What human-to-agent ratios are observed at different levels of practice, and how is responsibility distributed across the redefined roles? Qualitative studies of teams adopting the discipline are needed; grounded theory \citep{gonzalez-prieto_reliability_2021,diaz2021GT} is a natural methodological fit, and the constructs of Section~\ref{sec:patterns} (interaction patterns, competences) provide sensitizing concepts to be challenged by data. This line is not being opened here for the first time. The agile research community has been converging on it through a series of research--practice workshops, whose collaboratively produced roadmap already identifies team coordination, role evolution, and governance of AI teammates as priority concerns \citep{nilsxp}. What the present paper contributes to that agenda is a set of constructs, i.e., the interaction patterns and competences of Section~\ref{sec:patterns}, that can act as sensitizing concepts for the qualitative studies such a roadmap calls for.

\paragraph{RA2. Validation of the interaction patterns} How do the interaction patterns evolve across different organizational contexts and levels of agent autonomy? The five patterns of Section~\ref{sec:five-patterns} are a hypothesis. Mixed-methods studies combining interaction logs, observation, and surveys should test whether these are the right five, how they compose into workflows, and how they vary by domain and agent capability.

\paragraph{RA3. Development of orchestration competences} How do engineers acquire the competences of Section~\ref{sec:roles} (specification literacy, comparative reading, orchestration, norm articulation), which educational interventions accelerate them, and how should SE curricula incorporate specification literacy as a first-class skill? This line connects directly to the educational reimagining that \citet{hoda2026agentic} calls for.

\paragraph{RA4. Organizational adoption pathways} How do organizations adopt the proposed framework? This paper has deliberately not proposed an adoption or maturity model, since the cross-organizational evidence available to us is thin and largely self-reported. Multi-case studies should characterize how organizations sequence the mechanisms of Section~\ref{sec:harness}, which orderings succeed, where adoptions stall, and whether partial adoption is in fact worse than none, as the agentic model predicts.

\paragraph{RA5. Measurement frameworks} How should the proposed framework be measured? Existing team metrics (e.g., DORA) were designed for human-authored work. Constructs specific to this model require operationalization: specification quality, evidence-pack completeness, normative-specification maturity, consultation latency, and agent portfolio cost, including its energy dimension \citep{cruz2025innovating}.

Taken together, these research directions outline a program for transforming the proposed framework from a conceptual model into an empirically grounded engineering discipline. The framework introduced in this paper is intended not as a definitive methodology, but as a foundation upon which the software engineering community can progressively build, validate, refine, or refute its assumptions.

\section{Conclusion}\label{sec:conclusion}

The emergence of agentic software engineering represents more than the next step in GenAI-assisted programming. Authors argue that it requires a fundamental socio-technical reconfiguration in which specifications become the primary coordination artifact between humans and autonomous agents, and where governance, rather than model capability alone, determines whether increased individual productivity translates into improved team performance.

This paper proposes a coherent conceptual socio-technical framework for understanding that transition. Authors positioned Spec-Driven Development (SDD) as an enabling discipline for agentic software engineering and characterized the methodological harness that governs agent behavior at team scale. Authors further identified recurring human--agent interaction patterns and described the competences and organizational structures required to support them. Together, these contributions provide an integrated perspective that connects previously fragmented discussions across academic work, industrial practice, and emerging agentic development tools.

At the same time, the authors emphasize that the proposed framework should be understood as a conceptual socio-technical framework rather than as a validated engineering methodology. Because the phenomenon is still emerging and much of the available evidence originates from industrial practice and gray literature, the framework should primarily be viewed as a set of falsifiable hypotheses that organize current knowledge and motivate future empirical research.

Ultimately, if software engineering is entering an era in which autonomous agents become permanent members of development teams, then the central research challenge is no longer how to build better coding models, but how to design the socio-technical systems that allow humans and agents to collaborate effectively, safely, and at scale. Addressing this challenge is, in our view, the key to transforming individual AI productivity into sustainable improvements in team and organizational performance.

In this sense, the framework presented here is intended not as the final word on agentic software engineering, but as a foundation upon which the community can progressively build, validate, refine, and, where necessary, challenge its assumptions.

\begin{acks}
GenAI tools were used in a strictly assistive role to support English language editing and to suggest local revisions of the manuscript. All scientific content, methodological decisions, analysis and interpretation of results are the sole responsibility of the authors, who reviewed and validated every AI-assisted edit.

The authors also wish to thank practitioners such as MoureDev, \'Alvaro Moya, and Gentleman Programming, among others, for all the material they share on social media, from which the authors have learned a great deal.
\end{acks}

\section*{Declarations}
\textbf{Funding.} Not applicable. 

\textbf{Conflict of interest.} The authors declare that they have no conflict of interest.

\textbf{Data availability.} The replication package contains the source-selection material and extraction matrix used to support the synthesis reported in this paper. It is deposited in Zenodo under DOI \url{https://doi.org/10.5281/zenodo.22151221}.

\bibliographystyle{ACM-Reference-Format}
\bibliography{references}

\end{document}